%% file: main.tex
\documentclass[longauth]{aa}
\usepackage{silence}
\usepackage{graphicx}
\usepackage{natbib}
\usepackage{scalerel}
\usepackage{float}
\usepackage[table]{xcolor}
\usepackage{tabularx}

\usepackage{txfonts}
\usepackage[pdfencoding=auto,psdextra]{hyperref}
\hypersetup{
    colorlinks=true,
    linkcolor=blue,
    filecolor=magenta,      
    urlcolor=blue,
    citecolor=blue
}
\makeatletter
\renewcommand*\aa@pageof{, page \thepage{} of \pageref*{LastPage}}
\makeatother

\usepackage[utf8]{inputenc}

\usepackage[switch, modulo]{lineno}
\usepackage{dcolumn}
\newcolumntype{d}[1]{D{.}{.}{#1}}

\usepackage{euclid}
\usepackage{makecell}
\nolinenumbers
\let\linenumbers\relax
\begin{document}

\title{Euclid Quick Data Release (Q1)}
\subtitle{AstroVink: A vision transformer approach to find strong gravitational lens systems
\thanks{This publication has been made possible by the 
participation of more than a thousand volunteers in the 
Space Warps project. Their contributions are individually 
acknowledged at
\href{https://www.zooniverse.org/projects/aprajita/space-warps-esa-euclid/about/team}{https://www.zooniverse.org/projects/space-warps-esa-euclid/team}.}}

\include{authors}

\abstract{
We present \texttt{AstroVink}, a vision transformer classifier designed for efficient and automated identification of strong lens candidates in \Euclid imaging. We build upon the \texttt{DINOv2} encoder, fine-tuned to distinguish between lens and non-lens galaxies. Our base model, trained on simulated strong lens systems and labelled non-lenses, recovers 88 of the 110 lens candidates within the top 500 ranked candidates, corresponding to an inspection efficiency of one lens per 5.7 inspected objects in our test set. After the Q1 data release, which yielded about 500 lens candidates, we retrained the model using high-confidence lens candidates and new negatives, initially flagged as potential lenses by other classifiers but rejected during visual inspection. The retrained network further improves performance, achieving recovery of all 110 systems within the same ranking and reducing the inspection effort to one lens per 4.5 inspected objects, demonstrating that incorporating real examples significantly enhances model generalisation. An analysis of training subsets revealed that the inclusion of realistic negative examples played a key role in this improvement. Finally, we applied the retrained model to the full Q1 original selection of $\sim$1.08M targets, followed by a new round of Space Warps citizen-science inspection and expert vetting, where we identified a total of eight Grade~A and 26 Grade~B new lens candidates. These results demonstrate that transformer–based architectures can recover strong lens candidates with high efficiency in real \Euclid data, while substantially reducing the number of candidates requiring visual inspection.
}


\titlerunning{ }
\authorrunning{ }

\maketitle

\section{\label{sc:Intro}Introduction}
Strong lensing occurs when light from a background source is deflected by a massive foreground object, such as a galaxy. This can produce arcs, rings, or multiple images of the source. These lensing systems are powerful tools in astrophysics since they allow us to study the mass distributions in galaxies \citep{Gavazzi_2007, Nightingale_2019,Sonnenfeld2024,Shajib_2024}, observe magnified distant sources \citep{Welch2022}, and constrain cosmological parameters such as the Hubble constant \citep{Wong2019} and the properties of dark energy and dark matter \citep{vegetti2023stronggravitationallensingprobe,li2024}. However, these studies are often limited by the small number of confirmed strong lens systems, only a few hundred, a consequence of the intrinsic rarity of strong gravitational lensing. Increasing the sample size is essential to improve statistical precision and unlock new scientific insight \citep{Sonnenfeld_2021, Sonnenfeld_2022, Shajib_2024}. 

Since the early serendipitous discoveries \citep{Walsh1979}, lens searches have evolved from feature-based methods \citep{alard2006automateddetectiongravitationalarcs,Gavazzi_2014,Joseph_2014} to convolutional neural networks (CNNs; \citealp{LeCun89}), which have raised the number of known candidates from hundreds to over 15\,000 in recent years \citep{Jacobs_2017,Petrillo_2017, Petrillo_2018,Jacobs_2019,Jacobs2_2019,Li_2020,Cameras2021,Rojas2022,Savary_2022,Huang_2021,Nagam25,Q1-SP053,storfer2025gravitationallensesunionseuclid}. 
The first results of the Quick Data Release (Q1; \citealp{Q1cite}) of the \Euclid space telescope mission \citep{EuclidSkyOverview, Scaramella-EP1} have led to at least 500 new strong lens candidates using a combination of citizen science, machine-learning techniques and expert inspection \citep{Q1-SP048, Q1-SP052, Q1-SP053, Q1-SP054, Q1-SP059}. However, even the best performing CNN models could not recover all candidates without manually inspecting over 20\,000 targets, highlighting the limitations of current CNN-based pipelines.
More recently, a few lens finding studies have adopted vision transformer (ViT) encoders \citep{thuruthipilly2022finding, gonzalez2025discoveringstronggravitationallenses}, a newer type of artificial neural network (ANN) architecture \citep{ANNsOverview}. CNNs extract features using local convolutional kernels (small filters that detect simple patterns such as edges or textures) and build up more complex representations with each layer, where a layer refers to one processing stage of the network. In contrast to CNNs, ViTs process entire images as sequences of patches, treating the image as a set of regions rather than analysing a single region at a time. This enables the model to capture global relationships and long-range dependencies more effectively, as well as subtle features that are essential for distinguishing between classes.

In this work, we present {\tt AstroVink}\footnote{\url{https://github.com/SaamieVincken/AstroVink}}; an implementation of the \texttt{DINOv2} ViT framework \citep{oquab2024dinov2learningrobustvisual}. \texttt{DINOv2} has been trained on general large image data sets collected from the public domain -- not specifically including any astronomical data. 

The framework provides a family of ViT models that learn general purpose visual features via large scale self-supervised pre-training. This means they are trained on large collections of unlabelled images to learn generic visual features such as shapes, textures and spatial relationships. These models have proven effective across different scientific domains such as medical imaging \citep{baharoon2024evaluatinggeneralpurposevision, song2024generalpurposeimageencoder}, satellite remote sensing \citep{bou2024exploringrobustfeaturesfewshot}, and other areas within astronomy \citep{lastufka2025bridginggapexaminingvision}. These properties make \texttt{DINOv2} well suited to capture the extended and faint structures characteristic of gravitational lenses. In this work, we adopt the ViT-S/14 variant and fine-tune it for the task of identifying strong gravitational lenses in \Euclid data.

To establish a controlled baseline, we carried out a systematic series of experiments to determine an effective configuration for strong lens classification. These include testing variations in model initialisation, learning-rate, and input representation. A first version, a simulation-only baseline, is fine-tuned on the same \Euclid Q1 training sets as prior work (simulated lenses and non-lens galaxies). A second version is further fine-tuned with real Q1 lens candidates. All experiments for both networks are evaluated on a reserved test set constructed from real \Euclid Q1 data for fair comparison.

In this paper, Sect.~\ref{sc:data} describes the \Euclid data sets and input image preparation, including the creation of simulated and real training samples as well as the reserved test set. Section~\ref{sc:method} details the \texttt{DINOv2} ViT architecture, training configuration, and controlled setup. Section~\ref{sc:results} presents the results of the parameter tests and the performance of the simulation-only baseline. Section~\ref{sc:q1retraining} outlines the retraining with real Q1 candidates and its effect on lens recovery and false positives. Section~\ref{sc:newcandidates} summarises the additional candidates identified through visual inspection. Finally, Sect.~\ref{sc:discussion} provides a discussion, and Sect.~\ref{sc:conclusion} provides the conclusions.

\begin{figure}
    \centering    \includegraphics[width=\linewidth]{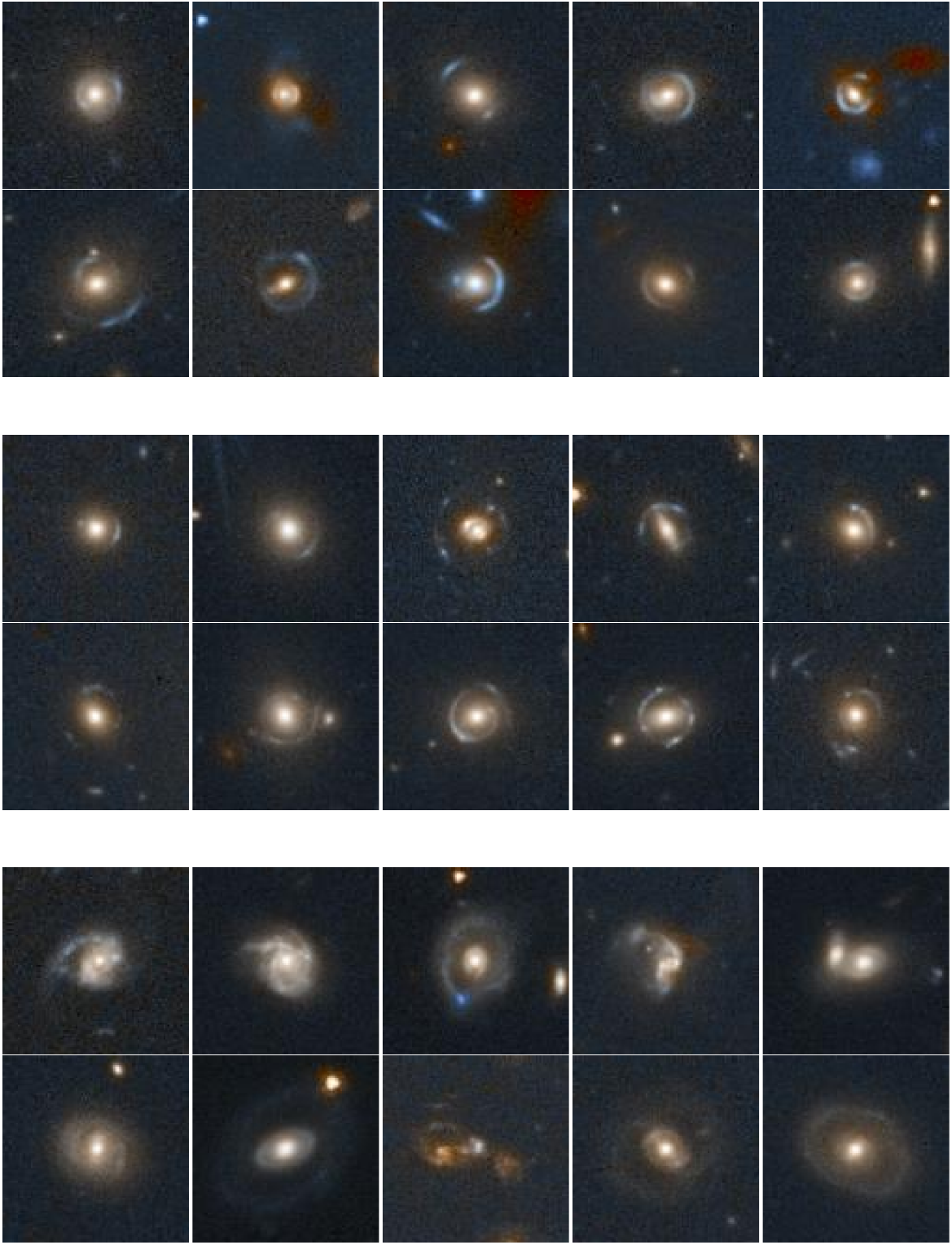}
    \caption{Examples of cutouts used during training and validation.
    The top group shows simulated strong lens systems, the middle group shows high-confidence lens systems identified in the Q1 data release, and the bottom group shows common false positives such as ring, spiral or merging galaxies. Each cutout has a size of $10\arcsec\times10\arcsec$ and is shown in \IE+\JE-MTF.}
    \label{fig:cutout-groups}
\end{figure}

\section{\label{sc:data}  Data } 
The \Euclid telescope provides imaging in one visible broad band, \IE, captured by the Visible Camera (VIS, \citealt{EuclidSkyVIS}), and three near-infrared bands, \YE, \JE, and \HE captured by the Near-Infrared Spectrometer and Photometer (NISP, \citealt{EuclidSkyNISP}). The images used in this work are generated from four combinations of the \Euclid photometric bands: \IE\ greyscale images; \IE+\YE\ and \IE+\JE\, two-band RGB composites (where the green channel is interpolated between the two bands); and a three-band \IE+\YE+\JE\ RGB mapping. Each cutout is generated using two different scaling methods: inverse hyperbolic sine function (hereafter arcsinh) with percentile clipping; and the Midtone Transfer Function (hereafter MTF). Both methods transform the pixel intensity distribution to enhance contrast. A detailed description of the eight different band and scaling combinations can be found in \cite{Q1-SP048}. MTF increases local contrast by compressing most pixels into a narrow intensity range (0.15--0.2), which makes arcs and galaxy features more defined. However, it also removes brightness differences, for example, a bright core and a faint arc may appear equally bright. Conversely, arcsinh preserves these differences by keeping faint regions faint and bright regions bright. All cutouts used in this work were created as JPEG (Joint Photographic Experts Group) files using the eight different image combinations; a representation of these combinations can be found in Figure 1 of \cite{Q1-SP053}. The original cutout size is $15\,\arcsecond{} \times 15\,\arcsecond{}$, which allows room for applying augmentation such as corner crops. After augmentation, the final cutout size used as input for the network is $10\,\arcsecond{} \times 10\,\arcsecond{}$. Throughout the development, we use a combination of simulated lens systems and common false positives, and later real lenses found in Q1. Some examples of these cutouts are displayed in Fig.~\ref{fig:cutout-groups}, and described in detail in the following subsections.

\subsection{\label{sc:q1-data-selection}Q1 data selection}
The \Euclid Q1 release provides space-based imaging across an area of $63\,\mathrm{deg}^{2}$, covering the three \Euclid Deep Fields: \Euclid Deep Field North, \Euclid Deep Field South, and \Euclid Deep Field Fornax \citep{Q1-TP001}. Inside this footprint, the Strong Lensing Discovery Engine (hereafter SLDE) selected a parent sample following the criteria described in \citet{Q1-SP048}. The selection criteria were designed to identify bright galaxies while removing stars and artefacts, resulting in approximately $10^{6}$ galaxies that potentially contain galaxy-galaxy strong lens candidates. We adopted this full parent catalogue for inference. The candidates in the SLDE follow a grading system that we also apply throughout this work, based on expert visual inspection (see \citealt{AcevedoBarroso24} for full details on the system). Grade~A lenses correspond to high-confidence systems, while Grade~B represent probable strong lenses where the confidence of experts is less certain. Grade~C corresponds to very low confidence systems, while Grade~X represents systems marked as a lens by previous ML approaches \citep{Q1-SP053} but classified as non-lens by the experts. In this work, we treat both Grade~A and B as lens systems of high significance. However, to treat a lens as confirmed it requires additional follow-up beyond imaging alone.

\subsection{\label{sc:ML_sets} Machine-learning sets}
ML models require large data sets to perform their tasks effectively, in this case, classifying lens systems (positives) among other non-lens galaxies (negatives). However, strong lensing is rare, and only a few lenses with available \Euclid data existed within the Q1 footprints at the time this project started. Adding to the difficulty, some of the negative classes, such as ring galaxies and mergers, are also rare, and we only have access to small samples of them. For the Q1 searches, a diverse set of realistic lens simulations and non-lens samples were used to train ML models. For our first model (simulation-only baseline) we used only the data described in \cite{Q1-SP053}. This allows a fair comparison with the previous models trained for the Q1 search. For our second model (retrained-model), we used the results from the searches already performed in Q1, including the catalogues of lenses and non-lenses compiled in the different papers. Here we summarise the data sets used to train and test the models presented in this work.

\subsubsection{\label{sc:simulated} Simulated lens systems}
We used targets from two sets of simulations, hereafter S1 and S2, following the same naming in \cite{Q1-SP053}. Both were generated by adding lensing features to real foreground galaxies. All simulations were provided in cutouts of $15\arcsec\times15\arcsec$ ($150\times150$ pixels).

S1 consists of simulations described in \cite{Rojas2022} and \cite{Q1-SP052}, generated using the {\tt Lenstronomy}\footnote{\url{https://github.com/lenstronomy/lenstronomy}} package \citep{Birrer_2018,Birrer_2021}. The deflectors are luminous red galaxies (LRGs) with known redshifts and velocity dispersions from the Early Data Release of the Dark Energy Spectroscopic Instrument (DESI-EDR; \citealp{DESIEDR2024}). Each deflector was modelled following a S\'ersic profile fitted to the \JE-band image to extract light-profile parameters for the mass model. Background sources are {\it Hubble}  Space Telescope (HST) F814W images with Hyper Suprime-Cam (HSC)-based colour information from \citet{Cameras2021}. To match \Euclid filters, the \IE\ band was approximated from HSC $r$ and $i$ bands, while the \YE-, \JE-, and \HE-bands were assigned by matching the $gri$ magnitudes of our sources to those in the COSMOS2020 catalogue \citep{Weaver_2022}, and adopting the corresponding VISTA $Y$, $J$, and $H$ magnitudes from the matched sources. Lens-source pairs were selected to yield Einstein radii larger than 0\farcs5. A singular isothermal ellipsoid (SIE) mass model was constructed using the light-profile parameters, velocity dispersion of the deflector, and redshifts of both lens and source. This model was then used to lens the background source light. The resulting images were downsampled to the \Euclid pixel scale, point-spread function (PSF)-convolved, and flux-scaled.

To augment the data, each deflector was rotated by $90^\circ$ increments and paired with a different source, creating four unique lensing configurations per deflector. These were grouped by rotation angle, with each subset containing approximately 2500 samples.

This set of simulations contains on the order of 11\,000 examples of lens systems. Additionally, we included an earlier version of the same kind of simulations, bringing in approximately 4000 more targets.

The S2 simulations were produced with the \texttt{GLAMER} lensing package \citep{Metcalf_2014, 10.1093/mnras/stu1860}. Cutouts of $200\times200$ pixels, centred on galaxies with apparent magnitude \IE < 22 were extracted. The selection excluded stars and nearly face-on spirals, but allowed a diverse morphological sample including elliptical and disc galaxies. Using a nearest neighbour algorithm, each target was matched to an object in the Flagship simulation \citep{EuclidSkyFlagship} based on the magnitudes in all four bands, ellipticity, and redshift. The parameters of the Flagship galaxy and dark matter halo were then used to construct a mass model for the lens. The simulation data used here were accessed via CosmoHub \citep{TALLADA2020100391,2017ehep.confE.488C}.

Source surface-brightness distributions were represented by one to four S\'ersic components. Total fluxes and effective radii were anchored to randomly selected HST Ultra-Deep Field galaxies at comparable redshifts \citep{Meneghetti_2008,Meneghetti_2010}. All sources were fixed at $z=5$, leaving variation in lens redshift, mass, and source properties to set the lensing diversity.

Light rays were traced through the composite mass model; lenses with Einstein radii below \ang{;;0.5} were discarded. The simulations were performed at four times the \IE-band resolution, then downsampled to \Euclid pixel scales before being convolved with the corresponding PSF. Synthetic lensed images were merged with the original survey frames, and cases with insufficient signal-to-noise or contrast relative to the lens galaxy were visually rejected. Further details are in Metcalf et al.\ (in prep.). In total, \num{5363} non-augmented images were generated.

Both sets combined gave us a total of approximately 23\,300 samples. All simulations (S1 and S2) were produced in all four \Euclid filters and transformed into the eight different colour-composite image combinations.

\subsubsection{\label{sc:non-lens} Non-lens galaxy samples}
The negative non-lens class includes general non-lens galaxies, and three known morphologies causing high false-positive rates \citep{Rojas2022} in lens finding; ring galaxies, spiral galaxies, and mergers.

For the negatives test set, we namely use the labelled targets from \cite{Q1-SP052}, consisting of approximately 2300 spirals, 250 mergers, 60 rings, 2300 other non-lens galaxies, and 2700 LRGs, the latter also serving as the base for the simulations in S1. This overlap ensures that the model is exposed to both lensed and non-lensed versions of similar deflector galaxies, helping it learn the distinction between genuine lensing features and intrinsic galaxy structure. In addition, a set of approximately 4000 randomly selected galaxies is generated during the creation of the S2 simulations but finally not used for lensing simulations. These cutouts were included as additional negative examples, bringing the total number of negative samples to approximately 11\,000.

\subsubsection{\label{sc:training_validation_sets}Training and validation sets}
The data sets we use in training and validation of the simulation-only baseline combine the simulated lenses of Sect.~\ref{sc:simulated} with the non-lens galaxies of Sect.~\ref{sc:non-lens}. Before any augmentation we divide the available $15\arcsec\times15\arcsec$ cutouts into a training pool (80\%) and a validation pool (20\%), resulting in approximately 13\,000 simulated lenses and 9300 non-lens for training and 3200 lenses and 2300 non-lenses for validation. This split at the catalogue level prevents leakage of nearly identical examples between the two pools.

To expose the network to positional and orientational variance, we generate augmentations of the original $15\arcsec\times15\arcsec$ cutouts. For both the positive and negative samples, we generate 14 derivatives of each original object. These include a $10\arcsec\times10\arcsec$ centre crop, eight non-overlapping $10\arcsec\times10\arcsec$ corner and edge crops (as illustrated in Appendix~\ref{fig:cropping-grid}), a horizontal flip, a vertical flip, and $90^{\circ}$, $180^{\circ}$ and $270^{\circ}$ rotations. For the low volume negative samples (merger and ring galaxies), additional flips were applied on top of the corner crops to further increase their representation. The resulting augmented training set contains approximately 182\,000 lens and 130\,000 non-lens images, while the validation set contains approximately 44\,000 lens and 32\,000 non-lens images. To avoid training bias, these sets were balanced by downsampling the lens samples to match the non-lens class. This balancing included filters to ensure no samples from low-volume classes like ring galaxies and mergers were removed. After balancing of the classes, we are left with a training set of 130\,000 and validation set of 32\,000 lens and non-lens samples. 

After the augmentation and balancing of data, we apply a final data leakage check. Data leakage occurs when information from outside the training data set, like from the validation set, is shared, or `leaked', with the model during training, leading to unreliable performance metrics. Since the simulated deflectors from S1 appear in four unique lensing configurations, the data leakage check is used to ensure no rotated or augmented cutouts from the same system or deflector are present in both the training and validation sets. This is done using perceptual hashing, a technique that represents each image by a numerical summary designed to preserve its visual appearance. Images with similar pixel-level structure are identified as similar using a 20\% pixel similarity threshold, following the method described by \cite{mckeown2023hamming}. The threshold is chosen to remove near-duplicate images arising from rotations or augmentations, while avoiding the removal of genuinely distinct systems. The check is applied to both lens and non-lens samples, resulting in four duplicate images to be removed. These four images are simply a duplicate entry of the same image but using a different image path which is why they were not flagged previously. Once the leakage check succeeds, it confirms that all validation samples are independent and not derived from the same base image as any of the training samples.

\subsubsection{\label{sc:q1-test-set}  Q1 test set }
After the Q1 release and the first lens finding results, a reserved test set was created to enable a controlled performance comparison between different neural networks trained on finding lens candidates. Throughout this work we will refer to this as the Q1 test set. This test set is excluded from any training or validation to ensure that later evaluations remain unbiased. The positive samples in the Q1 test set consist of 20\% of the combined Grade~A and B candidates from the Q1 SLDE catalogues (\citealp{Q1-SP048}; \citealp{Q1-SP052}; \citealp{EckerSLDE-F}).
Only high-confidence lenses were included in the test set, which resulted in 110 lens samples. The negative samples were drawn from the same catalogue. For this set, each object was either graded as non-lens by what is referred to as the Galaxy Judges (GJ) project, an internal \Euclid visual-inspection campaign where consortium members classified candidate systems, or excluded after initial rejection by Space Warps \citep[SW;][]{Q1-SP048}. SW is the \Euclid citizen science platform for strong lens discovery, where volunteers visually inspect \Euclid cutouts to identify features such as arcs, rings, or multiple images indicative of gravitational lensing. Each target is shown to multiple independent volunteers, and their classifications are aggregated to produce a consensus grade. From the total inspected non-lens targets, 75\% of a randomly selected set of 40\,000 was used for our negative set, which resulted in approximately 30\,000 non-lens cutouts for our Q1 test set. 

\section{\label{sc:method}  Method }
Identification of strong lens candidates in large amounts of \Euclid data requires a model that can reliably distinguish the characteristic features of lensing systems, such as rings, arcs, and multiple images surrounding a deflector, from non-lens systems with similar visual patterns. These potential sources of confusion include ring galaxies, spiral arms, tidal tails from mergers, and chance alignment of unrelated galaxies that can mimic lens-like configurations. Although CNNs have shown success in \Euclid galaxy-galaxy strong lens searches \citep{Q1-SP053}, their reliance on localised kernels often results in a high rate of false positives, motivating the use of a ViT architecture for this task.

\subsection{\label{sc:ViT}Vision transformer architecture}
The ViT used in this work builds on a mechanism originally introduced for natural language processing by \citet{vaswani2017attentionneed}, and later applied for image recognition by \citet{dosovitskiy2021imageworth16x16words}. Unlike CNNs, the ViT models each image as a sequence of fixed size, non-overlapping patches and uses a self-attention mechanism (Sect.~\ref{sc:attention-mechanism}). This design allows the network to learn global relationships between features in an image, like words in a sentence, needed to understand the correlation between background and foreground light sources or similar looking artifacts. 

We apply the mechanism by using the pre-trained ViT encoder referred to as \texttt{DINOv2} \citep{oquab2024dinov2learningrobustvisual}. Here, encoder refers to the component that analyses input images to extract visual features, and pre-training means that the encoder is trained beforehand on large collections of images so that it learns general visual features. The choice of the encoder is motivated by its ability to preserve extended information from faint and complex structures. Standard ViT architectures \citep{dosovitskiy2021imageworth16x16words, ruan2022visiontransformersstateart} represent an image as a sequence of vectors, referred to as tokens. Each token corresponds to an image patch and contains information about both the visual content of that patch and its position within the image.

One additional vector is the Classification (CLS) token, which represents a summary of the entire image. In most standard ViT implementations, this CLS token is used as the input for the final classification decision. In contrast, \texttt{DINOv2} combines the CLS token with the mean of all patch tokens, where this mean is obtained by averaging the patch representations so that all regions of the image contribute equally to the final summary. This is particularly relevant for gravitational lenses, where arcs and rings form a relationship between light sources in the image and fine details can be lost when compressed into a single token.

The \texttt{DINOv2} framework was released with a family of encoders, ranging from small to very large models (ViT-S/14, ViT-B/14, ViT-L/14, and ViT-g/14). In this work we adopt the ViT-S/14 variant, which contains 12 transformer layers, hereafter transformer blocks, and approximately 21 million parameters. This model provides sufficient capacity to capture the subtle and extended features of gravitational lenses, while remaining computationally efficient for training and inference on \Euclid scale data sets. Larger variants are significantly more demanding in terms of hardware and training resources, and did not provide any significant improvement in performance. 

All \texttt{DINOv2} models were pre-trained on the LVD-142M data set \citep{oquab2024dinov2learningrobustvisual}, a curated collection of about 142 million natural images. This data set was built from publicly available web-crawled sources, such as LAION \citep{schuhmann2022laion5bopenlargescaledataset}, but underwent extensive filtering to remove low-quality content, duplicates, and semantically inconsistent entries. The result is a high-quality, diverse image collection designed to provide strong and generalisable visual representations. No astronomical or \Euclid-like data were included in this pre-training stage, since the specific data set used is less important than its size and diversity. Large data sets expose the encoder to many different image conditions, such as brightness patterns and contrast variations, which improves its ability to adapt to new data during fine-tuning. 

The pre-training used a self-supervised distillation strategy. In this approach the network does not learn from labels, but instead improves by comparing different views of the same image. A `teacher' network, updated slowly during training, provides stable reference representations, and a `student' network is trained to reproduce them. This allows the model to learn useful features directly from the data without requiring explicit labels. This approach extends the original DINO method \citep{caron2021emergingproperties} by improving stability and scaling to very large data sets. It enables the model to learn transferable visual features, which are then, in this work, adapted into the domain of strong lens detection during fine-tuning.

\subsubsection{\label{sc:attention-mechanism}Self-attention mechanism}
The attention mechanism of a ViT determines how information from different parts of an image is combined, allowing the model to decide which regions of the image are most relevant when interpreting a given feature. For each patch, the model learns how strongly it should focus on every other patch in the image. This is done by converting each input vector into three components: a query $Q$, a key $K$, and a value $V$. These are learned linear transformations of the input; simple operations that allow the model to compare information between different image patches.

The attention weights are calculated by taking the dot product between the query and key vectors, scaled by the dimension of the key. These weights determine how much each patch should contribute to the output of another, and the resulting attention operation is

\begin{equation}
\text{Attention}(Q,K,V)=\operatorname{softmax}\!\left(\frac{QK^\mathsf{T}}{\sqrt{d_k}}\right)V~,
\end{equation}

\noindent where $d_k$ is the dimensionality of the key vectors. {Softmax\footnote{Softmax is a mathematical function that normalises the numbers produced by comparing one image patch to all other patches, converting them into positive values that sum to one.}} normalizes the scores, with the final output of the attention mechanism being a weighted sum of the value vectors. This full attention computation is applied to every patch in parallel, allowing the model to relate features across the full image. 

Using this mechanism we can create an attention map, which is a visual representation showing how strongly the ML model focuses on different regions in the image. This map is obtained by extracting the attention matrix from the final transformer block of the model. The attention matrix is a table with values that describe the attention assigned between image patches. The model computes several attention maps in parallel, which are averaged to obtain a single value per image patch. This map is reshaped into a 2D grid by arranging the patches according to their original positions in the image, and then resized to match the input image. The map can then get overlaid on the original images using a fixed colour map, to show how much attention the CLS token assigns to each patch when computing the final classification score. 

\subsection{\label{sc:kfold}K-fold cross-validation}
To assess how well the ViT generalises on different subsets of unseen data, we implement K-Fold cross-validation. This technique partitions the full data set into $k$ equally sized folds. In each iteration, one fold is used for validation while the remaining $k{-}1$ folds are used for training. This ensures that every image is used exactly once for validation and multiple times for training without overlap within a single fold. Splitting is done using shuffling and a {fixed random seed\footnote{A random seed is a fixed number used to initialise random number generation, so that the same random choices are made each time.}} for reproducibility. For each fold $i$, with a total of five folds, a performance metric $M_i$ is calculated. These individual fold scores are then averaged to obtain the final cross-validation score denoted by $\hat{M}$ in

\begin{equation}
\hat{M} = \frac{1}{k} \sum_{i=1}^{k} M_i~,
\end{equation}

\noindent where $M_i$ is the performance metric for the $i$-th fold and $k$ is the total number of folds. By averaging performance metrics across all folds, we obtain more insights on the expected performance over various subsets of data. This is particularly relevant for upcoming \Euclid data releases, where the statistical properties of the data may vary between sky regions or observation periods, and the available labelled samples may not fully represent future data. 

\subsection{\label{sc:training}Training configuration}
The training process adapts the \texttt{DINOv2} ViT-S/14 encoder to the specific challenges of strong lens classification. Without proper tuning, neural networks can mistakenly treat noise or unrelated features as lensing signals, even if they do not correspond to a real lens. This problem is known as overfitting, and each part of the training process is chosen to reduce this risk and improve the network's ability to generalise on new data.

Training begins with preprocessing to match the input format of the ViT encoder. Since the \texttt{DINOv2} weights {weights\footnote{Weights are numerical parameters that determine how input data is interpreted and how it contributes to its final outputs.}} were learned on three-channel RGB images, each arcsinh-\IE cutout, originally single-channel (greyscale), is stacked into three identical channels. Images of $100\times100$ pixels are then resized to $224\times224$, and pixel values are rescaled from 0--255 to 0--1 to improve numerical stability. Finally, the mean and standard deviation ($\mu=\{0.485, 0.456, 0.406\}$, $\sigma=\{0.229, 0.224, 0.225\}$) are applied for normalization. These values are derived from ImageNet, a large general data set commonly used to train and benchmark neural networks, and are consistent with the normalization used in the original \texttt{DINOv2 setup.}

Before entering the ViT, the image is divided into non-overlapping $14 \times 14$ pixel patches, resulting in a sequence of 256 patch embeddings. Here, an embedding refers to a numerical vector that represents the visual information contained in a patch. Each patch embedding is then linearly projected, meaning it is transformed into a fixed-length vector. The collection of these vectors is referred to as the embedding space, which is a numerical coordinate system in which all patches are represented and serves as the input for the transformer blocks. To preserve spatial information about where each patch came from in the original image, learnable positional encodings are added to each embedding. Here, the embeddings represent what is in each patch, and the positional encodings represent where that patch came from in the image. The CLS token is also added to the sequence and acts as an extra element that collects information from all patches through self-attention (detailed in Sect.~\ref{sc:attention-mechanism}). As the sequence passes through the transformer blocks, the CLS token is updated repeatedly and gradually becomes a compact summary of the entire image.

To make the training set more varied and limit the risk of overfitting, data augmentation is applied before patching. Each image can be flipped horizontally and vertically with a 50 percent chance. These flips do not change the features of the lens but provide additional versions of each example, helping the model learn that a lens can appear in any orientation.

After passing through all transformer blocks, the CLS token is extracted and averaged with the mean of all patch embeddings. Combining both global information from the CLS token and distributed information from the patches helps the model capture extended arcs or faint structures that might span multiple patches. 

Before projection, the vector undergoes layer normalisation; a technique used to adjusts the distribution of the input features by centring them around zero. This normalisation improves numerical stability and makes the following blocks easier to train. This choice is consistent with the \texttt{DINOv2} ViT-S/14 encoder itself, which already uses layer normalisation inside every transformer block to stabilise representations. The final combined vector containing all features goes to a classification head, which turns it into two raw scores. 

The classification head on top of the encoder is made up of three fully connected layers {fully connected layers\footnote{A fully connected layer, also called a dense layer, is a neural network layer in which every output unit is connected to every input unit from the previous layer, allowing information from all inputs to be considered simultaneously.}}. Between these layers the model uses an activation function called GELU (Gaussian error linear unit). An activation function introduces non-linearity, allowing the network to learn relationships beyond simple straight lines. GELU does this in a smooth way rather than with abrupt steps, which helps transformers capture subtle differences in the data. The classification head includes dropout with a rate of 0.1 applied after the first activation, which means that during training, the network randomly sets 10 percent of the intermediate values to zero. This forces the network to rely on multiple features instead of depending too much on any single one. This setup is consistent with the pre-trained \texttt{DINOv2} transformer blocks themselves, which also contain dropout for regularisation.

During training, the raw class scores produced by the classifier are compared to the true labels using the standard cross-entropy (CE) loss-function. This loss-function is defined as

\begin{equation}
\label{eq:ce}
\mathcal{L}_{\mathrm{CE}} = -\ln(p_y)~,
\end{equation}

\noindent where \(p_y\) is the predicted probability for the correct class \(y\). This penalises the model based on the confidence assigned to the true label. A loss-function measures how far the predicted probabilities are from the correct answer, a low loss meaning good separation between the two classes (lens and non-lens), and forms the direction of adjusting the network's weights. To update these weights, the training uses an optimiser. An optimiser is the algorithm that changes the network’s adjusting the network step by step to reduce the loss. We adapt a version of the widely used Adam optimiser \citep{kingma2017adammethodstochasticoptimization}, called \texttt{\texttt{AdamW}}, which was introduced by \cite{loshchilov2019decoupledweightdecayregularization}. This adaptation includes weight decay, which discourages the weights from becoming too large. This helps the network to avoid learning extreme or unstable values that could lead to overfitting. 

The final output is a two-element vector of probabilities, $\big(p_{\mathrm{lens}},\, p_{\mathrm{non\mathchar`-lens}}\big)$, each ranging from 0 to 1. These scores represent the likelihood that the input image belongs to the `lens' or `non-lens' class, respectively. Finally Softmax is applied to the raw output logits to normalise the scores, ensuring they can be interpreted as valid probabilities. 

\subsection{\label{sc:controlledsetup}Controlled setup}
Controlled experiments test how individual training settings affect lens recovery and ensure that the ViTs performance is reproducible. These tests isolate the impact of input representation, loss-function, and training configuration on lens recovery and false positive rates. Settings chosen before training, such as the {learning-rate\footnote{The learning-rate controls the size of the updates applied to a networks' weights during training.}}. or loss-function, are referred to as hyperparameters. All experiments use the same reserved Q1 test set (see Sect.~\ref{sc:q1-test-set}), ensuring that changes in performance come from the tested settings and not from differences in the data.

Model performance is evaluated using three complementary measures: the Receiver Operating Characteristic (ROC) curve, the Area Under the Curve (AUC), and lens recovery within the top $N$ ranked candidates. A ROC curve  shows the classifier’s performance by plotting the true positive rate (the fraction of lenses correctly identified) against the false positive rate (the fraction of non-lenses incorrectly classified as lenses) as the decision threshold is varied. The AUC is the Area Under this Curve and provides a single scalar measure of class separability, where a value of 1 corresponds to perfect separation and 0.5 corresponds to random classification. Lens recovery within the top $N$ ranked candidates measures how many known lenses are found when inspecting only the $N$ highest scoring objects. Here, top $N$ refers to objects ranked by the model’s predicted lens-likelihood score $\big(p_{\mathrm{lens}})$, in descending order.

All controlled experiments are ran on a single Graphics processing unit (GPU) without parallelisation, as an additional measure to avoid variability from parallel computation. The hardware configuration consists of an NVIDIA RTX A4500 with CUDA version 12.8, a single CPU core, and 16\,GB of reserved RAM. Data loading is performed using a single process (worker) to avoid variability introduced by parallel data loading. Within the \texttt{AdamW} optimiser, we apply a weight decay value of 0.1, which means that during training an extra penalty equal to 0.1 times the sum of the squared weights is added to the loss \citep{loshchilov2019decoupledweightdecayregularization}. This factor controls how strongly the optimiser discourages large weights. This value is consistent with the range used for ViTs in transfer learning \citep{dosovitskiy2021imageworth16x16words, oquab2024dinov2learningrobustvisual}. 

The learning-rate for both the encoder and classifier is set to $5 \times 10^{-6}$ as a starting point, with an alternative picked up after the experiment described in Sect.~\ref{sc:learning-rate-seed}. During training, the learning-rate follows a cosine annealing schedule with warm restarts \citep{loshchilov2017sgdrstochasticgradientdescent}. The learning-rate at epoch \(t\) follows

\begin{equation}
\eta_t = \eta_{\min} + \frac{1}{2}(\eta_{\max} - \eta_{\min})
\left[1 + \cos\!\left(\frac{T_{\mathrm{cur}}}{T_i}\pi\right)\right]~,
\end{equation}

\noindent where \(\eta_{\max}\) is the initial learning-rate, \(\eta_{\min}\) is the lower bound, \(T_{\mathrm{cur}}\) is the number of epochs since the last restart, and \(T_i\) is the cycle length. When \(T_{\mathrm{cur}}=T_i\), the learning-rate reaches \(\eta_{\min}\); when \(T_{\mathrm{cur}}=0\) immediately after a restart, it resets to \(\eta_{\max}\). In this work we set \(T_0=5\) (first restart after five epochs) and \(T_{\mathrm{mult}}=2\) (doubling the cycle length after each restart).

Training is performed in mixed precision, a technique where lower numerical precision is used for some operations, to reduce memory use without losing accuracy. Before each update step, gradient norm clipping is applied to cap the total gradient size, preventing unstable weight changes. To confirm the reliability of the setup, additional runs over various learning-rate and random seed values were performed (detailed in Sect.~\ref{sc:learning-rate-seed}).

For all experiments, we apply the arcsinh-\IE scaling and photometric band combination, since the following tests are not designed to compare input representations. Full performance comparisons across bands are presented in Sect.~\ref{sc:band_scaling_section}.

\section{\label{sc:results} Results}
This section presents the results of optimizing the network’s hyperparameters and input configurations to identify the best training setup for strong lens classification on \Euclid data. We carried out a series of controlled experiments to examine the influence of random seed selection, learning-rate settings, and input image choices (photometric band and scaling combinations). Each experiment was designed to test one factor at a time while keeping all other conditions the same so that the impact of that single factor on performance could be clearly seen.

\subsection{\label{sc:learning-rate-seed} Seed and learning-rate configuration}
Random seed selection affects the initialization of model weights and the random processes during training, including data shuffling, dropout operations, and weight initialization. To measure how much variation there was and ensure reproducible results, we tested 10 different random seeds using the \IE-arcsinh combination with a fixed learning-rate of $5 \times 10^{-6}$ for both encoder and classifier.

Besides seed selection, optimizing learning-rates requires careful consideration of the two component architecture: the pre-trained \texttt{DINOv2} encoder and the randomly initialized classification head. The encoder, having been pre-trained on natural images, requires a lower learning-rate to preserve learned representations while allowing fine-tuning for astronomical features. The classifier, being randomly initialized, can accommodate higher learning-rates for faster convergence. We tested nine combinations of encoder learning-rates ($1 \times 10^{-6}$, $2 \times 10^{-6}$, $5 \times 10^{-6}$) and classifier learning-rates ($1 \times 10^{-5}$, $2 \times 10^{-5}$, $5 \times 10^{-5}$) using the \IE-arcsinh input configuration. The numerical results of these tests are detailed in Appendix~\ref{table:lr_config}.

The corresponding visual comparison of the tests are detailed in Appendix~\ref{fig:lr-seed-combined}. The shaded $\pm1\sigma$ regions illustrate variation across seeds (blue) and learning-rate configurations (orange). The best performance was achieved using $5 \times 10^{-6}$ for both encoder and classifier, confirming that moderate learning-rates for both components provide the most effective balance between preserving pre-trained features and enabling adaptation to the data set. We fixed this configuration (seed = 1, learning-rates = $5 \times 10^{-6}$) for all subsequent experiments due to its reproducibility and performance. In contrast, the highest tested rate ($2 \times 10^{-5}$) produced the poorest performance, demonstrating that overly aggressive fine-tuning degrades performance.

\subsection{\label{sc:band_scaling_section} Photometric band and scaling comparison}
One key question is which of the eight photometric band and scaling image combinations is most effective for identifying strong lens candidates. While hyperparameter optimisation was carried out using the \IE-arcsinh input, it remained important to test whether this choice also represented the optimal input for the final model configuration. To do this, we applied the best-performing model setup identified earlier to all available input combinations: \IE, \IE+\JE, \IE+\YE, and \IE+\YE+\JE, each prepared with arcsinh and MTF scaling.

Eight variations of the simulation-only baseline were trained, each using the same objects and training parameters but a different band-scaling input. Each variation was evaluated on the Q1 test set (see Sect.~\ref{sc:q1-test-set}) prepared with the same representation.

The results are described by a ROC curve shown in Fig.~\ref{fig:band-scaling}, the corresponding AUC values are reported in Appendix~\ref{table:band_scaling}. The \IE-arcsinh input produced the best ROC curve and has the highest AUC of 0.983, while \IE+\YE+\JE-MTF yielded the lowest, at 0.752. Configurations based on \IE-band data consistently achieved higher scores than those dominated by the \YE- and \JE-band data. Across nearly all configurations, arcsinh scaling produced higher AUC scores than MTF.

These results indicate that \IE-arcsinh is the strongest input representation for this model. However, they do not exclude the possibility that colour information could contribute to model performance in other configurations. The experiment shows a preference for \IE-arcsinh under the current training setup, however it is important to note that the JPEG format does not preserve the full instrumental information, and the RGB channels cannot be directly mapped to the VIS and NISP bands. A more detailed evaluation of multi-band input would require training and testing on images where each channel is explicitly aligned with a corresponding spectral band. Such an investigation lies beyond the scope of the present study. Accordingly, the comparison reported here reflects relative behaviour under the JPEG-based Q1 setup and should not be interpreted as a definitive assessment of the contribution of the NISP bands.

\begin{figure}
\centering
\includegraphics[width=\linewidth]{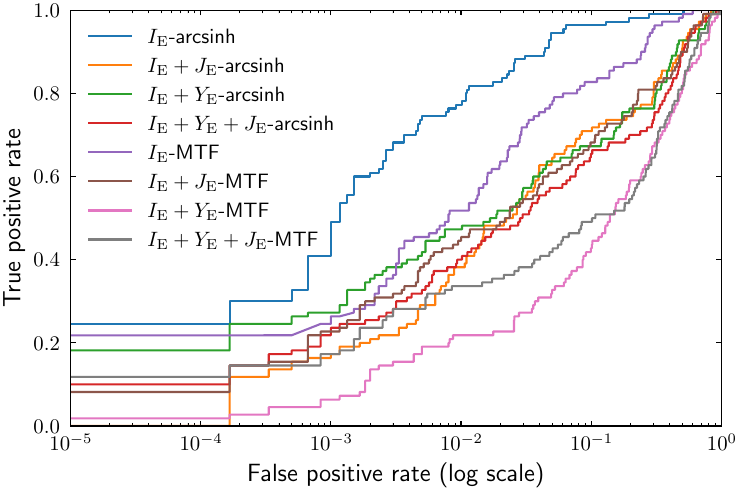}
\caption{ROC curve comparison of all eight combinations of VIS (\IE) and NISP (\YE, \JE) bands with arcsinh and MTF scaling. The x-axis shows the false positive rate (logarithmic scale), defined as the fraction of non-lens systems incorrectly classified as a lens. The y-axis shows the true positive rate, defined as the fraction of correctly identified lens systems. Each curve shows one variation of the \texttt{\texttt{AstroVink-base}} model trained with a different band and scaling configuration. The curves show the relative performance of each input representation; \IE-arcsinh achieves the highest performance, while \IE+\YE+\JE-MTF performs worst.}
\label{fig:band-scaling}
\end{figure}

\subsection{\label{sc:Best_ML_results} Best model results}
After testing the effects of seed selection, learning-rates, loss-function, and input band-scaling combinations, we identified the configuration that produced the strongest results on the Q1 test set. The final model, here after \texttt{AstroVink-base}, is a vision transformer trained with the \texttt{AdamW} optimizer, cross‑entropy loss, and a cosine annealing learning-rate scheduler, with encoder and classifier learning-rates of $5\times10^{-6}$. 

All encoder blocks were unfrozen during training to allow full adaptation to \Euclid data. For training, we input \IE-arcsinh images in batches of 32. We set up 200 epochs for training, and to avoid overfitting, we applied early stopping with a patience of 20 epochs based on the lowest validation loss, although \texttt{AstroVink-base} ended up needing only seven epochs across the data set before it reached a plateau. The average runtime per epoch was approximately nine minutes, with the total runtime just over an hour. 

\texttt{AstroVink-base} achieved an AUC of 0.983 on the Q1 test set. This indicates a high {true positive rate\footnote{The true positive rate is the fraction of positive examples that the model correctly classifies as positive.}} across different thresholds. In order to understand how the network interprets the data, we used the self‑attention mechanism of the vision transformer to visualise the final block of the network (as described in Sect.~\ref{sc:attention-mechanism}). The result is the attention map displaying which regions are most important for the model’s final prediction.

We applied \texttt{AstroVink-base} to a sample of Q1 targets to obtain a score ($P_{\mathrm{lens}}$) and build for each an attention map. These targets all received a label using the Q1 grading scheme (as described in Sect.~\ref{sc:q1-data-selection}). This allowed us to examine whether the network’s internal focus changes systematically between clear lenses, uncertain cases, and non‑lenses. 

\begin{figure*}
\centering
\includegraphics[width=\textwidth]{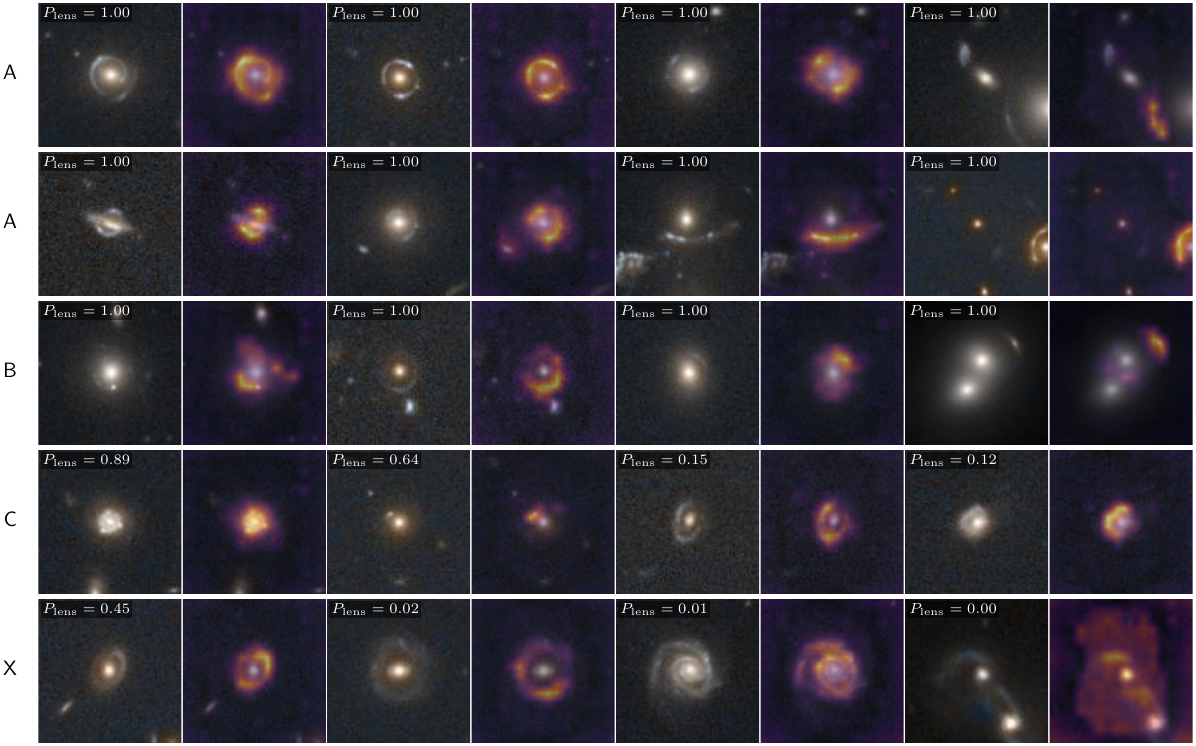}
\caption{Attention maps from the final transformer block for Q1 cutouts. Each pair of panels shows a single galaxy cutout (left) and the corresponding attention map created by \texttt{AstroVink-base} (right) overlaid on the original input cutout. The attention maps are overlaid on the input images using a fixed colour scale, where brighter colours indicate regions that receive higher attention from the model when computing the final classification score. 
Rows are grouped by graded targets where Grade~A and B show high-confidence strong lens candidates, Grade~C shows low-confidence candidates, and Grade~X shows non-lens systems. The value ($P_{\mathrm{lens}}$) of each image, is the lens probability given by the network. Some examples include lenses that are not centred within the cutout, yet the model still assigns high $P_{\mathrm{lens}}$ values, illustrating that it does not rely on strict central positioning.}
\label{fig:attention-mechanism}
\end{figure*}

The results are shown in Fig.~\ref{fig:attention-mechanism}. The attention maps predominantly highlight curved and extended structures, such as lensed arcs, spiral arms, and rings, independent of their position within the image. This indicates that such structures play a central role in determining the final classification score.

While the maps highlight similar curved structures in both lens and non-lens systems, the distinction between them is reflected in the assigned lens-probability. In particular, non-lens systems with lens-like morphologies receive attention on these features, but are assigned a low score. This indicates that the network recognises these structures as relevant, but can still distinguish subtle morphological differences between lenses and non-lenses. Targets that do not contain clear lensing structures often receive low scores and exhibit attention that is distributed across the entire image rather than concentrated on specific features.

For the K-fold cross validation (as described in Sect.~\ref{sc:kfold}), the metrics F1-score, precision, and recall are used to analyse how well the classifier distinguishes lenses from non-lenses. Precision measures how many of the images predicted as lenses are actually lenses, while recall measures how many of the true lenses are correctly identified. The F1-score is the harmonic mean of precision and recall and provides a single value to balance both aspects. Analysing these metrics for each fold shows how consistently the model ranks lens candidates highly across the entire data set. Table~\ref{table:kfold_results} gives the results of the validation. The results show consistent performance across all folds, with mean metrics all around 0.98 indicating stable generalisation behaviour.

\begin{table}
    \centering
    \caption{
    Results from 5-fold cross-validation of the \texttt{AstroVink-base} classification model. The first column lists the fold index, where each fold corresponds to one partition of the data. The second column reports the precision, the third column reports the recall, and the fourth column reports the F1-score. The final two rows show the mean value and the standard deviation of each metric across all folds, indicating the overall performance level and its variability due to the choice of validation split.}
    \smallskip
    \label{table:kfold_results}
    \smallskip
    \begin{tabular}{l@{\hskip 3em}c@{\hskip 3em}c@{\hskip 3em}c}
        \hline\hline
        Fold & Precision & Recall & F1-score \\
        \hline
        1 & 0.983 & 0.980 & 0.980 \\
        2 & 0.980 & 0.982 & 0.981 \\
        3 & 0.988 & 0.993 & 0.990 \\
        4 & 0.993 & 0.986 & 0.989 \\
        5 & 0.988 & 0.992 & 0.989 \\
        \hline
        Mean & 0.986 & 0.987 & 0.986 \\
        Std  & 0.005 & 0.006 & 0.005 \\
        \hline
    \end{tabular}
\end{table}

\section{\label{sc:q1retraining} Q1 retraining}
The catalogue created in Q1 includes 250 Grade A and 247 Grade B candidates from the SLDE, representing high-confidence lenses. Retraining of the network used a subset of the Q1 data that was fully separated from the Q1 test set. The sample comprised 380 high-confidence candidates and 4726 non-lenses. Here the non-lenses are systems flagged as potential lenses by Q1 networks but rejected after both citizen science and expert inspection, making them valuable samples for training. Throughout this work, we refer to these objects as hard-negatives: systems that are confirmed non-lenses but closely resemble genuine strong lens systems in their visual and morphological properties, making them challenging negative examples for automated classifiers. These hard-negatives should not be confused with false-negatives, which would correspond to true lenses incorrectly classified as non-lenses. The purpose of retraining is to adapt a model initially trained on simulated data (Sect.~\ref{sc:simulated}) to the real \Euclid domain since the diverse range of real lens systems is not fully captured by simulations. 

\subsection{\label{sc:retraining_method} Q1 retraining method}
We first retrained the model following the same configuration as \texttt{AstroVink-base}. All transformer blocks were unfrozen to allow full adaptation to the new data, the \texttt{\texttt{AdamW}} optimiser with CE loss was used together with the same learning-rate schedule and batch size, and training was again performed on \IE-arcsinh inputs. However, an unexpected drop in performance was observed: the network found fewer lenses in the Q1 test set than when trained only on simulations.

A closer inspection showed that the retrained network failed to recognise several simulated-like lens systems it had previously detected, while also not generalising well to the new Q1 examples. This behaviour indicated that the previously learned representations had been overwritten during retraining, pointing to catastrophic forgetting. Catastrophic forgetting refers to the loss of previously learned information when a model is trained on new data. The effect was amplified by a domain shift between simulated and real \Euclid images: simulations provide controlled examples, but their brightness, noise, and structural patterns might not fully match the diversity of systems in Q1. As a result, the network lost generalised features learned from simulations while still failing to capture the full variability of the real data.

To mitigate this, retraining was carried out in stages. All simulated data from the base training were combined with the available Q1 candidates, and then gradually replaced: in each round, 10\% of the simulations were removed and replaced with 10\% of the Q1 examples, until only the Q1 candidates remained. This progressive blending preserved useful features from the simulations while adapting the model to the real survey domain. Catastrophic forgetting was further prevented by freezing earlier encoder blocks when appropriate and by adopting a new loss-function designed to handle imbalance and focus learning on harder examples, as will be detailed in the following sections.

\subsubsection{\label{sc:freezing_strategy}Freezing strategy}
Fine-tuning all pre-trained blocks of a transformer is one of the reasons a network can suffer from catastrophic forgetting. The earlier blocks of the encoder mainly capture generic patterns such as edges and textures, while the later blocks adapt to domain-specific information. To preserve the features learned from simulations, the earlier blocks were kept fixed during retraining, while only the later blocks were adjusted on the Q1 data. The point at which to separate fixed and trainable transformer blocks was determined using two analyses: CLS token probing; and centred kernel alignment (CKA; \citealt{kornblith2019similarityneuralnetworkrepresentations}). 

\begin{figure}
    \centering  \includegraphics[width=\columnwidth,height=0.75\columnwidth,keepaspectratio]{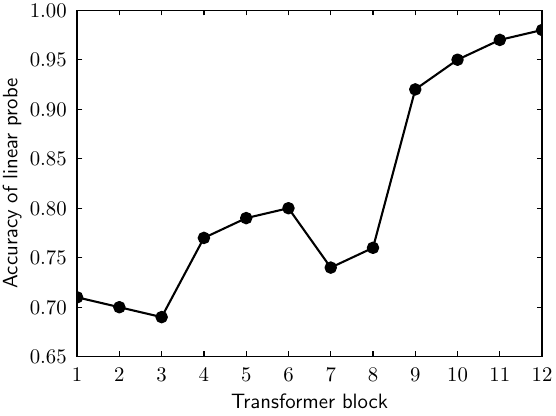}
    \caption{Block-wise probing of the CLS token output. The x-axis shows the transformer block index. The y-axis shows the validation accuracy, the fraction of correctly classified samples, of the linear probe test per block. For each transformer block, the encoder was frozen and a linear classifier was trained on the CLS representation to assess separability between lenses and non-lenses. Validation accuracy remains below 0.8 through block\,8, rises sharply to 0.92 at block\,9, and exceeds 0.95 in blocks\,10--12. This shows that discriminative information only becomes well defined in the final third of the encoder, with a transition around block\,9.}
    \label{fig:linear-probe}
\end{figure}

The CLS probe tests whether class-relevant information is already encoded at a given block. For this, the encoder is frozen and a classification head is trained on the CLS token from a single block. The validation accuracy then reflects how easily lenses and non-lenses can be separated using that representation. As shown in Fig.~\ref{fig:linear-probe}, accuracy remains below 0.8 through block\,8, then increases sharply to 0.92 at block\,9 corresponding to a relative increase of 15\% compared to earlier blocks and exceeds 0.95 in blocks\,10--12. This demonstrates that discriminative information only becomes well defined in the final third of the encoder, with a clear transition around block\,9.  

\begin{figure}
    \centering  \includegraphics[width=\columnwidth,height=0.75\columnwidth,keepaspectratio]{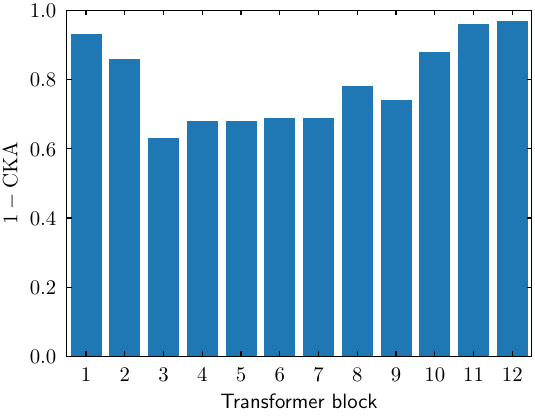}
    \caption{Representation shift during retraining, measured as $1-\mathrm{CKA}$ similarity between base encoder (before fine-tuning on Q1 data) and fine-tuned encoder activations for each transformer block. The x-axis shows the transformer block index. The y-axis shows $1-\mathrm{CKA}$, where CKA measures the similarity between feature representations before and after fine-tuning. Higher values correspond to stronger changes, while lower values indicate greater similarity. Blocks~1--2 shift substantially, blocks~3--9 remain relatively stable, and blocks~10--12 exhibit the largest changes. This indicates that early blocks retain general low-level features, while the final transformer blocks adapt strongly to the real \Euclid data.}
    \label{fig:cka-shift}
\end{figure}

The second analysis quantifies how much each block changes during fine-tuning. Here, the representational similarity between the base and fine-tuned models is measured using CKA. This provides a normalised similarity score between 0 and 1, where 1 indicates identical representations and lower values correspond to greater changes in the learned features. Figure~\ref{fig:cka-shift} shows $1-\mathrm{CKA}$ for each block: blocks\,1--2 shift substantially; blocks\,3--9 remain relatively stable; and blocks\,10--12 exhibit the strongest changes. This indicates that early blocks retain general low-level features, while the final transformer blocks adapt strongly to the real \Euclid data. Block\,9 shows intermediate behaviour, its CLS accuracy increases sharply but its representation shifts less than block\,8, suggesting that it already encodes useful discriminative information that is refined further during retraining.  

Although blocks 1--2 show relatively large shifts in the CKA analysis, the CLS probe indicates that they do not yet encode discriminative information. Their changes therefore reflect low-level adjustments rather than useful class separation. For this reason, they are also kept frozen during retraining. Based on the combined interpretation, blocks\,1--8 are frozen and blocks\,9--12 are updated.

\subsubsection{\label{sc:loss_comparison}Loss function}
In \texttt{AstroVink-base} training, standard CE loss was used, since the data set was relatively balanced and consisted entirely of clean simulated images. Under those conditions, CE performed well, and no major issues were observed in classification performance.

This changed during retraining on \Euclid data. The new data set was highly imbalanced, with far fewer true lenses than non-lenses, and many of the non-lenses were difficult false positives objects. This introduced ambiguity that was not present in the simulations. In this setting, CE loss began to underperform. Since it treats all examples equally, the gradient is dominated by the majority class. The model can reduce its total loss by confidently predicting obvious non-lenses, while failing to adjust its predictions on rarer cases.

To address this, focal loss \citep{lin2018focallossdenseobject} was introduced. This loss-function modifies CE by reducing the impact of well classified examples and focusing training on those that the model misclassified. This is especially useful in imbalanced data sets, where improving performance on a specific minority class is more important than minimising the average loss. The focal loss-function 

\begin{equation}
\label{eq:focal}
\mathcal{L}_{\text{focal}} = -\alpha_y (1 - p_y)^\gamma \ln(p_y)~,
\end{equation}

\noindent uses \(p_y\) as the predicted probability for the correct class \(y\), \(\alpha_y\) is a class-specific weight, and \(\gamma > 0\) is the focusing parameter. The term \((1 - p_y)^\gamma\) reduces the contribution of correctly classified examples (\(p_y \rightarrow 1\)) to the loss. 

When \(\gamma = 0\), the equation becomes equivalent to standard CE. The \(\alpha\) term allows control over class imbalance, while \(\gamma\) adjusts how strongly the model concentrates on uncertain predictions. The parameters \(\alpha\) and \(\gamma\) are set as fixed constants when initialising the loss-function and are not updated during training. Specifically, we set $\alpha = 2.0$ for lenses, $\alpha = 1.0$ for non-lenses, and $\gamma = 1.0$. For strong lens detection, where true lenses are rare and many non-lenses are visually similar, this set-up helps the network focus on the most informative and difficult examples.

\subsection{\label{sc:q1results} Q1 retraining results}
The retraining strategy was applied to adapt the ViT to the \Euclid Q1 domain. The \texttt{AstroVink-Q1} was trained on progressively mixed data sets, starting from simulated images and gradually incorporating real \Euclid cutouts until only Q1 examples remained. Following the results from the CLS and CKA analyses, part of the encoder blocks were kept frozen during training.

The model was optimised using \texttt{AdamW} with a cosine-annealing learning-rate schedule, differential learning-rates of \(1\times10^{-5}\) for the encoder and \(5\times10^{-5}\) for the classifier, and a batch size of~32. Training employed focal loss to handle class imbalance and focus learning on the most informative examples. Early stopping with a patience of~20 epochs was used to retain the best-performing checkpoint. Each retraining round used the progressively updated data set; consequently, the number of epochs and convergence time varied per stage, with the complete staged retraining taking approximately 2.5 hours on a single GPU.

\begin{figure}
    \centering
    \includegraphics[width=\columnwidth]{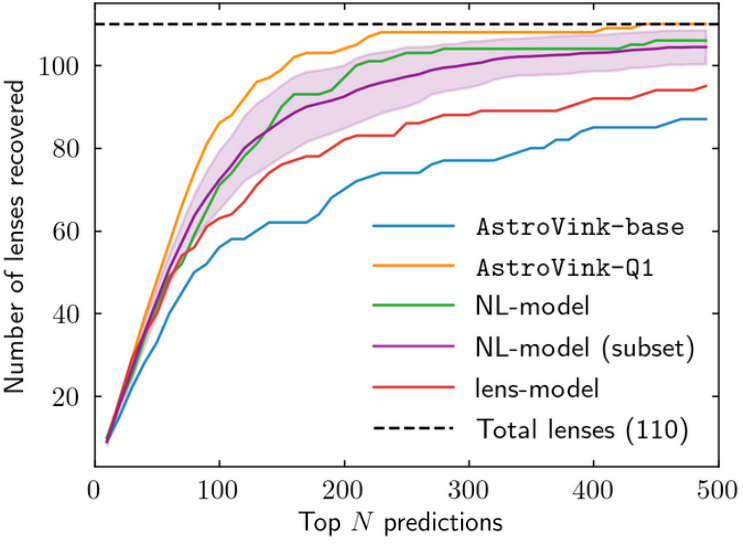}
    \caption{Recovery of known Q1 lenses as a function of the top $N$ ranked candidates for different retraining configurations. The x-axis shows the top $N$ predictions, referring to the highest-ranked objects based on the network’s predicted lens-likelihood score. The y-axis shows the number of lenses from the Q1 test set that the network recovered within the top $N$. The blue curve shows \texttt{AstroVink-base} trained only on simulations. Orange corresponds to \texttt{AstroVink-Q1} trained on the full Q1 retraining set (lenses + non-lenses). The red and green curves isolate the effect of using only the Q1 lens (lens-model) or only the Q1 non-lens (NL-model) sets. The purple curve shows the non-lens configuration where the number of non-lenses was matched to the 380 available lenses using ten different random subsets; the purple shaded band indicates the $\pm 1\sigma$ range across these runs. The dashed line marks the total number of known lenses in the Q1 test set (110).}
    \label{fig:lens-non-lens-contribution}
\end{figure}

Figure~\ref{fig:lens-non-lens-contribution} summarises the performance of the retrained models on the Q1 test set. The results show the number of known lenses recovered as a function of the top $N$ ranked candidates, where candidates are ordered by the networks' predicted lens-likelihood score. 

\texttt{AstroVink-Q1} (orange curve) shows to be the best performing network, trained on the full retraining set in addition to the original simulations. This model represents the maximum recovery capability achieved in the present work, recovering 86 of the 110 known lenses within the top 100 candidates and 109 within the top 300 of the Q1 test set .

With respect to \texttt{AstroVink-base} (blue curve), the retrained network achieves a substantial improvement in lens recovery. The gain is most pronounced in the top few hundred candidates, where the prioritisation of true lenses over contaminants is most impactful for follow-up visual inspection. The combined use of real positive and hard-negative Q1 examples enables the network to better suppress morphologically similar non-lenses, such as mergers and ring galaxies, without compromising recall of genuine lenses.

We further investigated the individual impact of the Q1 subsets in order to understand if real lenses or hard-negatives played a more important role in the retraining. To assess this, we followed the same retraining strategy but created two different sets, set-lens; with only the Q1 lens subset in combination with negative examples from \texttt{AstroVink-base}; and set-negatives; with only the Q1 non-lens subset in combination with the original simulation set. As shown in the figure, all Q1-based variants outperformed the simulation-only baseline, but with clear differences between subsets.

The model retrained on the set-lens data set, hereafter `lens-model', improved recovery compared to the base model (90 versus 88 lenses within the top 500 predictions, corresponding to 5.5 versus 5.7 objects to be inspected to find one lens), but its performance remained significantly limited compared to the final Q1-retrained model, which combines the full data set and achieved complete recovery of all 110 lenses (4.5 inspections per lens) in the top 500. For set-negatives, we first retrained the model using the entire negative data set (approximately 4700 examples) together with approximately 4000 simulated lenses from the base training. This configuration, hereafter `NL-model', achieved a higher recovery than the lens-model retrained on set-lens alone, indicating that the addition of hard-negatives improved discrimination. However, one could argue that the large number of negatives compared to real lenses introduced a bias in favour of the non-lens class, partly explaining the stronger performance of the NL-model.
  
To address this, a second test was conducted in which we retrained 10 independent models, each using a set composed of 380 non-lenses. The ten non-lens subsets were selected randomly to avoid bias from any particular sample. In this case, the shaded band around the purple line represents the variance across these ten runs. This experiment confirmed that the benefit of adding hard-negatives is stable across selections, while showing that the observed improvement was not simply caused by the larger training set size but by the higher information content of the negative examples themselves. 

The combination of both subsets delivered the best overall performance, recovering the largest number of lenses across all $N$. This demonstrates that exposure to both real positive examples and hard-negative non-lenses is essential for optimal classification performance in \Euclid-scale searches. All numerical results of Fig.~\ref{fig:lens-non-lens-contribution} are detailed in Appendix~\ref{tab:q1-retraining-results}. 

\section{\label{sc:newcandidates} Visual inspection and additional candidates}
The original Q1 search in 1.08 million cutouts over 63~deg$^2$ yielded 497 Grade A and B lenses from the main discovery engine catalogue \citep{Q1-SP048}. An extension catalogue was later published by \cite{EckerSLDE-F}, presenting 72 additional strong lenses missed in the initial search due to a bias against bright, low-redshift systems. This set includes 38 Grade~A and 34 Grade~B candidates, increasing the Q1 sample by over 10\% and adding systems of particular interest, such as edge-on discs, red sources, and a double-source-plane candidate. 

We applied \texttt{AstroVink-Q1} to the same cutouts used in the original Q1 search. The objective was to identify highly scored systems absent from any of the original Q1 catalogues, thereby recovering strong lens candidates missed during the initial discovery effort.

All cutouts in the Q1 parent sample were assigned a lens probability score by \texttt{AstroVink-Q1} and sorted in descending order. Any object that had been previously shown to volunteers during the Q1 inspections, regardless of its classification outcome, was removed from the list. This ensured that the resulting candidate set consisted solely of systems never before inspected by volunteers.

For this project, as in the original Q1 search \citep{Q1-SP048}, we made use of the SW platform. We took the top 10\,000 highest ranked objects from the network. After cross-matching with previously inspected objects on the platform, including all targets part of the SLDE catalogue, we were left with a total of approximately 6300 uninspected candidates. For each target, four image combinations were prepared: \IE-only; \IE+\YE; \IE+\JE; and \IE+\YE+\JE. The \IE-only combination matched the input representation used during network training, while the additional variants provided complementary morphological and colour information to aid in classification. From the 6300 inspected candidates, 907 were voted as potential strong lens candidates.

Following the citizen science phase, a second round of inspections was conducted by experts in the strong lensing domain. Each of the 907 candidate systems was independently graded by multiple experts, using the same grading (A, B, C, and X) scheme as in the Q1 inspections. A target was only retired from the workflow after receiving at least ten independent expert classifications.

The individual grades were then mapped to numerical values (X:0, C:1, B:2, A/A+:3) as defined in \citet{Q1-SP048}. The final score was calculated as the average of the assigned grade values. This helps us to sort the targets from more to less likely to be a lens candidate. 

To separate the candidates into final grades (A, B, C, X), we decided on thresholds by displaying the candidates sorted by descending scores. This ensures that targets are a faithful representation of their grades. The adopted cutoffs were 2.5 for Grade\,A, 2.0 for Grade\,B, and 1.4 for Grade\,C, with any lower values assigned to Grade~X (non-lens). 

Applying these cutoffs yielded a total of nine Grade~A, 27 Grade~B, and 72 Grade~C lens candidates. One grade~A and seven grade~C targets were previously reported by \cite{AgileLensXi}, and one grade~B was previously reported by \cite{EckerSLDE-F}, leaving a total of eight Grade~A, 26 Grade~B, and 65 Grade~C of totally new systems in the Q1 footprint. The list of Grade~A and Grade~B candidates is provided in Table~\ref{table:new_AB_targets}, and a mosaic of these newly discovered systems is shown in Appendix~\ref{fig:new-targets-AB}. Additionally, the catalogue with new candidates is published on {Zenodo}\footnote{\url{https://zenodo.org/records/17425610}}.

\section{\label{sc:discussion}  Discussion}
When we compare our results to other ML approaches applied to Q1 \citep{Q1-SP053}, where the best performing model identified 164 Grade~A/B candidates within its top 1000 ranked objects, we find that the simulations-only \texttt{AstroVink-base} model identified 235 Grade~A/B candidates in its top 1000.

To further analyse the performance we investigate the inspection efficiency of each \texttt{AstroVink} network, defined as the average number of inspected candidates required to recover one lens candidate from our Q1 test set. \texttt{AstroVink-base} recovers 88 of the 110 confirmed lenses within the top 500 ranked candidates, corresponding to an inspection efficiency of one lens per 5.7 inspected objects. After retraining with real \Euclid lens and non-lens cutouts, \texttt{AstroVink-Q1} achieves complete recovery (110 / 110) and improves the inspection efficiency to one lens per 4.5 inspected objects. 

We note that there are statistical uncertainties associated with the composition of the Q1 test set, since the number of confirmed lenses remains limited and the sample of inspected non-lenses is not fully representative of the true distribution. Nevertheless, the relative model behaviour on this set provides meaningful insight. 

We applied our Q1-retrained-network, \texttt{AstroVink-Q1}, to the original Q1 set to identify any additional systems missed in the original catalogue. This search gave us an additional 36 high-confidence systems, of which two systems overlap with other searches.

A preliminary characterisation of the newly identified systems, shown in Appendix~\ref{fig:new-targets-AB}, reveals several edge-on lenses that were absent from the original Q1 catalogues. This likely reflects the fact that the initial models were trained only on simulations, which at the time did not include such configurations. Incorporating real \Euclid data during retraining therefore improves the network’s ability to recognise rarer morphologies, highlighting its importance for uncovering lens populations that are under-represented or absent in simulated training sets.

\section{\label{sc:conclusion}  Conclusion}
We have trained and evaluated a vision transformer for strong lens detection using \Euclid Q1 imaging. All experiments used the same reserved Q1 test set, which contains 110 confirmed lenses and about 30\,000 non-lenses selected from the Q1 SLDE catalogue. With this setup, differences in results reflect model choices and not the data. 

The best configuration of the network was trained with \texttt{AdamW} and a cosine schedule using equal learning-rates of \(5\times10^{-6}\) for encoder and classifier. Repeated training with different random seeds confirmed that the setup is stable. A systematic comparison of eight input representations showed that cutouts in \IE-arcsinh scaling gave the best results with an AUC score of 0.983, while the weaker options like \IE+\YE+\JE-MTF dropped to an AUC score of 0.752. Additionally we validated the network across five validation folds where the model achieved mean F1, precision, and recall close to 0.986, showing stable and reproducible performance. 

By combining high-confidence lenses from the \Euclid domain with realistic and difficult non-lens samples, the retrained network \texttt{AstroVink-Q1} reached an AUC of 0.994 on the Q1 test set and recovered 109 of the 110 true lenses within the top 300 highest ranked images. This concentration of lenses indicates that transformer-based classifiers enable large-scale lens discovery with \Euclid by reducing the number of cutouts requiring human inspection

Attention map inspections show that the classifier consistently attends to arc-like structures, both in genuine lenses and in look-alike systems such as ring galaxies, spirals, or mergers. The difference lies in the score where lenses are ranked highly, while non-lenses with similar shapes receive low confidence. This confirms that the model has learned to separate true lensing features from common false positives.

In summary, our work delivers a controlled, reproducible, and highly efficient lens ranking method. It reduces the volume of cutouts needing inspection while recovering nearly all known lenses within a small top-ranked set. The classifier provides a promising outlook for strong-lens science in the forthcoming \Euclid data releases.

\paragraph{Code availability.}
The \texttt{AstroVink} source code and documentation are available at
\url{https://github.com/saamievincken/AstroVink}.
The repository contains the model architecture, inference scripts, and usage instructions.
The trained weights and data used in this work are based on internal \Euclid processing
and are not publicly released.

%
%

\begin{acknowledgements}
\AckQone
\AckEC  
\AckCosmoHub
\end{acknowledgements}
 
%
%

\bibliography{Euclid, Q1, my} 

%

\begin{appendix}
\onecolumn

\section{Data augmentations}
To increase training diversity and reduce overfitting, each $15\arcsec\times15\arcsec$ cutout was augmented through systematic crops and rotations. 
This appendix illustrates the crop layout used to generate the $10\arcsec\times10\arcsec$ inputs for model training.

\begin{figure*}
    \centering
    \includegraphics[width=\textwidth]{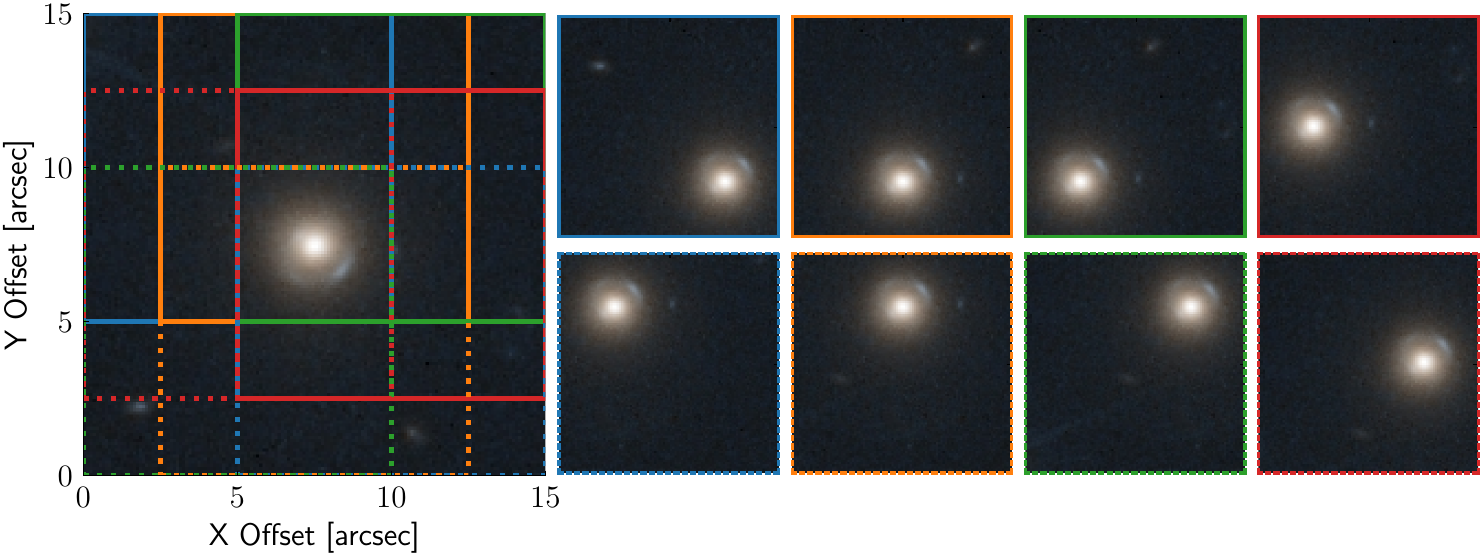}
    \caption{Visual layout of the corner and edge crops extracted from a $15\arcsec\times15\arcsec$ training cutout. 
    The original image (left) shows the crop regions, each highlighted by a coloured box, and the eight corresponding $10\arcsec\times10\arcsec$ crops (right). 
    Solid outlines indicate corner crops, while dotted outlines mark edge crops.}
    \label{fig:cropping-grid}
\end{figure*}

\section{Band and scaling comparison}
This appendix summarises the controlled experiments performed to assess the influence of input band combinations and scaling methods on model performance. The ROC curves shown in Fig.~\ref{fig:band-scaling} and the corresponding AUC values reported in Table~\ref{table:band_scaling} quantify the relative performance of eight input representations tested with the \texttt{AstroVink-base} model on the Q1 validation set.

\begin{table}
\centering
\caption{Area under the ROC curve (AUC) for different input representations tested with the \texttt{AstroVink-base} model on the Q1 validation set. Each configuration combines VIS (\IE) and NISP (\YE, \JE) bands with either arcsinh or MTF scaling.}
\smallskip
\label{table:band_scaling}
\smallskip
\begin{tabular}{l@{\hskip 2em}c}
\hline\hline
Input representation & AUC \\
\hline
$I_\mathrm{E}$-arcsinh & 0.983 \\
$I_\mathrm{E}+J_\mathrm{E}$-arcsinh & 0.868 \\
$I_\mathrm{E}+Y_\mathrm{E}$-arcsinh & 0.861 \\
$I_\mathrm{E}+Y_\mathrm{E}+J_\mathrm{E}$-arcsinh & 0.840 \\
$I_\mathrm{E}$-MTF & 0.941 \\
$I_\mathrm{E}+J_\mathrm{E}$-MTF & 0.864 \\
$I_\mathrm{E}+Y_\mathrm{E}$-MTF & 0.752 \\
$I_\mathrm{E}+Y_\mathrm{E}+J_\mathrm{E}$-MTF & 0.772 \\
\hline
\end{tabular}
\end{table}

\section{Learning-rate and seed test}
\begin{figure}
\centering
\includegraphics[width=0.7\columnwidth]{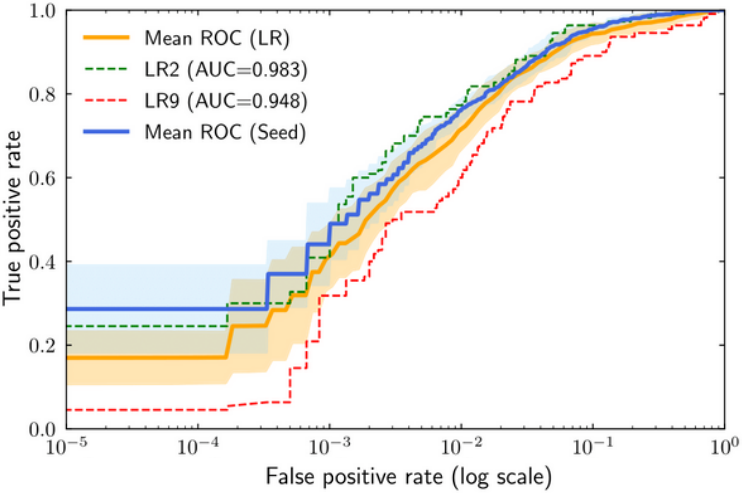}
\caption{Combined ROC curves for seed variation and learning-rate experiment using \texttt{AstroVink-base}. The x-axis shows the false positive rate on a logarithmic scale, corresponding to the fraction of non-lens systems incorrectly ranked as lenses above a given threshold. The y-axis shows the true positive rate, corresponding to the fraction of recovered lens systems. The blue shaded area shows the $\pm1\sigma$ range across ten random seeds, and the orange shaded area shows the $\pm1\sigma$ range across nine learning-rate configurations. The best and worst performing learning-rate configurations are marked with dashed lines, and include their corresponding AUC score in the legend.}
\label{fig:lr-seed-combined}
\end{figure}

This appendix summarises the controlled experiments performed to assess the influence of random seed selection and learning-rate configuration on model stability. 
The curves in Fig.~\ref{fig:lr-seed-combined} and the values in Table~\ref{table:lr_config} quantify the variability across ten seeds and nine learning-rate pairs tested on the Q1 validation set. In the table, the first column  lists the configuration ID; the second column reports the learning-rate for the encoder; the third column reports the learning-rate for the classifier; and the fourth column reports the AUC score based on the ROC curve in Appendix~\ref{fig:lr-seed-combined}. We see that both LR2 and LR5 achieve best performance with an AUC score of 0.983, while LR9 achieves worst performance with an AUC of 0.948.

\begin{table}
\centering
\caption{Tested learning-rate configurations for the encoder and classifier of \texttt{AstroVink-base}, with the resulting performance.}
\smallskip
\label{table:lr_config}
\smallskip
\begin{tabular}{l@{\hskip 2em}c@{\hskip 2em}c@{\hskip 2em}c}
\hline\hline
ID & Encoder-LR & Classifier-LR & AUC \\
\hline
LR1 & $2\times10^{-6}$ & $2\times10^{-6}$ & 0.973 \\
LR2 & $5\times10^{-6}$ & $5\times10^{-6}$ & 0.983 \\
LR3 & $5\times10^{-6}$ & $1\times10^{-5}$ & 0.982 \\
LR4 & $1\times10^{-5}$ & $1\times10^{-5}$ & 0.955 \\
LR5 & $2\times10^{-6}$ & $1\times10^{-5}$ & 0.983 \\
LR6 & $1\times10^{-5}$ & $2\times10^{-6}$ & 0.980 \\
LR7 & $1\times10^{-6}$ & $5\times10^{-6}$ & 0.966 \\
LR8 & $5\times10^{-6}$ & $2\times10^{-6}$ & 0.979 \\
LR9 & $2\times10^{-5}$ & $2\times10^{-5}$ & 0.948 \\
\hline
\end{tabular}
\end{table}

\section{Recovery statistics}
This appendix lists the recovery of known Q1 lenses within the top $N$ ranked candidates for each training configuration. The performance curves in Fig.~\ref{fig:lens-non-lens-contribution} and the numerical results in Table~\ref{tab:q1-retraining-results} quantify these results. The evaluated networks are as follows: \texttt{AstroVink-base} was trained only on simulations; \texttt{AstroVink-Q1} on the full Q1 retraining set (lenses + non-lenses); lens-model on Q1 lenses only; NL-model on Q1 non-lenses only; and NL-model (subset) on ten random non-lens subsets matched in size to the lens sample. In the table, each column reports the cumulative number of recovered lenses within the top 100, 300, and 500 highest ranked candidates based on the lens-likelihood score given by each network. The total number of lenses in the Q1 test set is 110.

\begin{table}
    \centering
    \caption{Number of known Q1 lenses recovered within the top $N$ ranked candidates for each retraining configuration shown in Fig.~\ref{fig:lens-non-lens-contribution}.}
    \label{tab:q1-retraining-results}
    \begin{tabular}{lccc}
    \hline\hline
    Training set & Top 100 & Top 300 & Top 500 \\
    \hline
    \texttt{AstroVink-base} & 56 & 77 & 88 \\
    \texttt{AstroVink-Q1} & 86 & 109 & 110 \\
    lens-model & 63 & 88 & 90 \\
    NL-model & 76 & 104 & 107 \\
    NL-model (subset) & 77 & 100 & 104 \\
    \hline
    \end{tabular}
\end{table}

\section{New targets}
This section displays the mosaic of newly identified strong lens candidates recovered by the Q1-retrained vision transformer model. The examples shown here correspond to systems graded A and B during the expert inspection.

\begin{figure*}
    \centering
    \includegraphics[width=\textwidth]{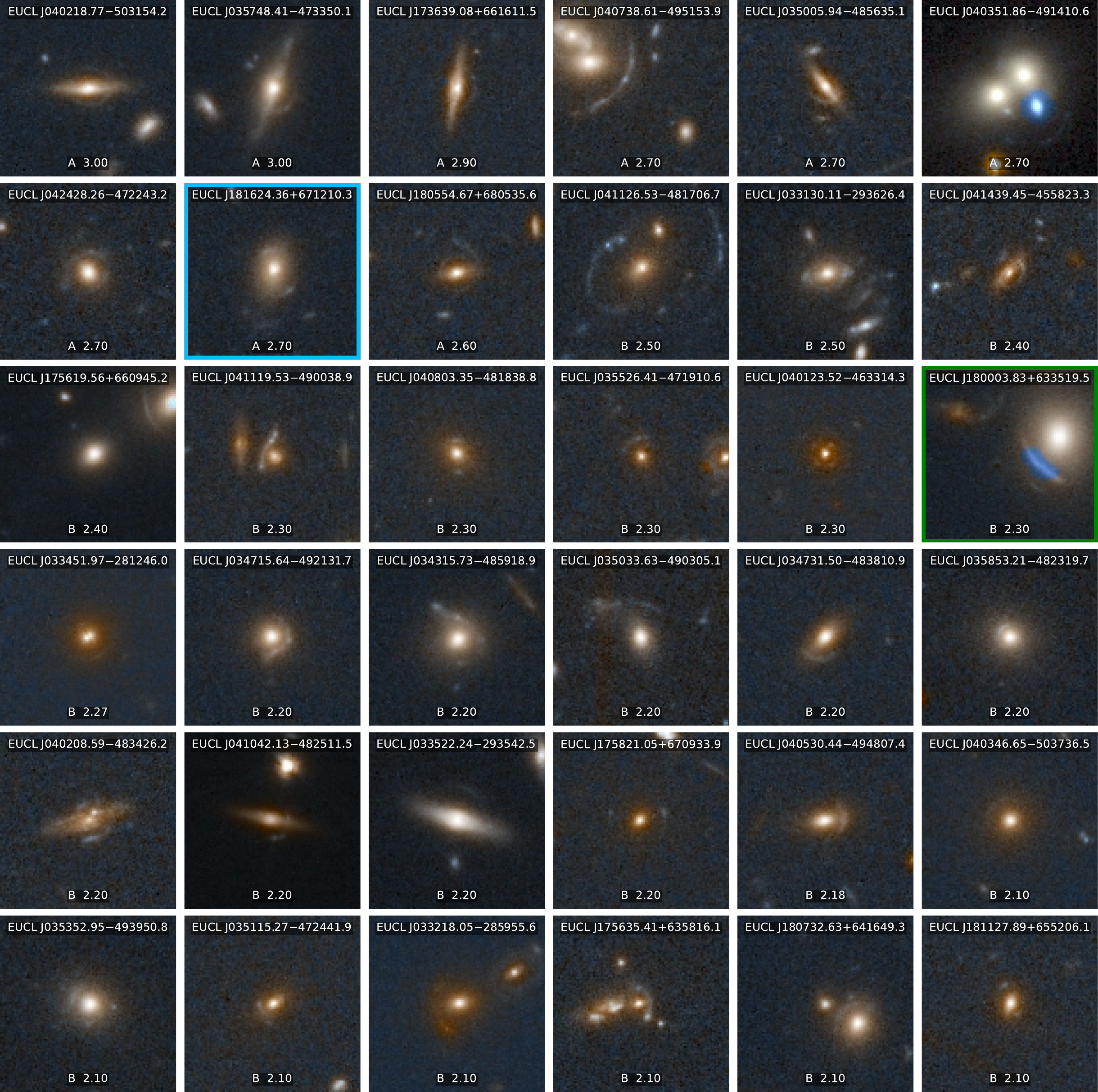}
    \caption{
    Newly identified Grade~A and Grade~B lens candidates recovered by \texttt{AstroVink-Q1} shown in \IE+\JE-MTF. The overlapping candidate from \cite{AgileLensXi} is marked with a blue frame (EUCL\,J181624.36$+$671210.3). The overlapping candidate from \cite{EckerSLDE-F} is marked with a green frame (EUCL\,J180003.83$+$633519.5). Each cutout is annotated at the top with its identifier, derived from right ascension and declination of the MER detection. The label at the bottom indicates the final expert grade and the visual inspection score reflecting confidence in the lens classification. Some Grade~A candidates appear edge-on. These systems were largely absent in the simulated training data used for the initial Q1 searches, and additional examples are now recovered after retraining \texttt{AstroVink} with real Q1 examples that include these morphologies. We explicitly show some cutouts in which the target is off centred to demonstrate that \texttt{AstroVink} is able to detect such configurations. However, for catalogue efficiency and ease of reuse we report the coordinates and object name of the centred version of the target.}
    \label{fig:new-targets-AB}
\end{figure*}

\begin{table}
\centering
\caption{Grade A and B lens candidates identified in this work and overlapping with prior literature. The first column reports the object name. The second and third columns report the right ascension (RA) and declination (Dec) in decimal degrees. The fourth column reports the averaged expert visual inspection score ($\mathrm{VI}_{\mathrm{score}}$), the fifth column reports the assigned confidence grade, and the sixth column reports the source of the original identification. The table is ordered by RA descending. }
\smallskip
\label{table:new_AB_targets}
\smallskip
\begin{tabular}{l d{5} d{5} c c l}
\hline\hline
Name & RA & Dec & VI\_score & Grade & Discovery \\
\hline 
EUCL\,J181624.36$+$671210.3 & 274.10151 & 67.20289 & 2.70 & A & [1] \\
EUCL\,J180554.67$+$680535.6 & 271.47780 & 68.09323 & 2.60 & A & This work \\
EUCL\,J173639.08$+$661611.5 & 264.16284 & 66.26988 & 2.90 & A & This work \\
EUCL\,J042428.26$-$472243.2 & 66.11776 & -47.37869 & 2.70 & A & This work \\
EUCL\,J040738.61$-$495153.9 & 61.91090 & -49.86499 & 2.70 & A & This work \\
EUCL\,J040351.86$-$491410.6 & 60.96612 & -49.23630 & 2.70 & A & This work \\
EUCL\,J040218.77$-$503154.2 & 60.57821 & -50.53172 & 3.00 & A & This work \\
EUCL\,J035748.41$-$473350.1 & 59.45174 & -47.56394 & 3.00 & A & This work \\
EUCL\,J035005.94$-$485635.1 & 57.52477 & -48.94309 & 2.70 & A & This work \\
EUCL\,J181127.89$+$655206.1 & 272.86624 & 65.86837 & 2.10 & B & This work \\
EUCL\,J180732.63$+$641649.3 & 271.88596 & 64.28037 & 2.10 & B & This work \\
EUCL\,J180003.83$+$633519.5 & 270.01596 & 63.58875 & 2.30 & B & [2] \\
EUCL\,J175635.41$+$635816.1 & 269.14756 & 63.97115 & 2.10 & B & This work \\
EUCL\,J175821.05$+$670933.9 & 269.58773 & 67.15944 & 2.20 & B & This work \\
EUCL\,J175619.56$+$660945.2 & 269.08152 & 66.16257 & 2.40 & B & This work \\
EUCL\,J041439.45$-$455823.3 & 63.66441 & -45.97315 & 2.40 & B & This work \\
EUCL\,J041126.53$-$481706.7 & 62.86054 & -48.28521 & 2.50 & B & This work \\
EUCL\,J041119.53$-$490038.9 & 62.83140 & -49.01081 & 2.30 & B & This work \\
EUCL\,J041042.13$-$482511.5 & 62.67555 & -48.41988 & 2.20 & B & This work \\
EUCL\,J040803.35$-$481838.8 & 62.01396 & -48.31078 & 2.30 & B & This work \\
EUCL\,J040530.44$-$494807.4 & 61.37685 & -49.80207 & 2.18 & B & This work \\
EUCL\,J040346.65$-$503736.5 & 60.94439 & -50.62681 & 2.10 & B & This work \\
EUCL\,J040208.59$-$483426.2 & 60.53583 & -48.57395 & 2.20 & B & This work \\
EUCL\,J040123.52$-$463314.3 & 60.34801 & -46.55398 & 2.30 & B & This work \\
EUCL\,J035853.21$-$482319.7 & 59.72175 & -48.38883 & 2.20 & B & This work \\
EUCL\,J035526.41$-$471910.6 & 58.86005 & -47.31963 & 2.30 & B & This work \\
EUCL\,J035352.95$-$493950.8 & 58.47063 & -49.66413 & 2.10 & B & This work \\
EUCL\,J035115.27$-$472441.9 & 57.81366 & -47.41164 & 2.10 & B & This work \\
EUCL\,J035033.63$-$490305.1 & 57.64016 & -49.05144 & 2.20 & B & This work \\
EUCL\,J034731.50$-$483810.9 & 56.88128 & -48.63638 & 2.20 & B & This work \\
EUCL\,J034715.64$-$492131.7 & 56.81517 & -49.35882 & 2.20 & B & This work \\
EUCL\,J034315.73$-$485918.9 & 55.81557 & -48.98861 & 2.20 & B & This work \\
EUCL\,J033522.24$-$293542.5 & 53.84268 & -29.59516 & 2.20 & B & This work \\
EUCL\,J033451.97$-$281246.0 & 53.71655 & -28.21280 & 2.27 & B & This work \\
EUCL\,J033218.05$-$285955.6 & 53.07524 & -28.99879 & 2.10 & B & This work \\
EUCL\,J033130.11$-$293626.4 & 52.87547 & -29.60734 & 2.50 & B & This work \\
\hline
\end{tabular}
\smallskip

\textbf{References:}
\hspace{1em}[1] \cite{AgileLensXi}, [2] \cite{EckerSLDE-F}
\end{table}

\end{appendix}

\end{document}

%% file: authors.tex
\newcommand{\orcid}[1]{} 
\author{Euclid Collaboration: S.~H.~Vincken\orcid{0009-0005-7305-2359}\thanks{\email{saamie.vincken@fhnw.ch}}\inst{\ref{aff1}}
\and K.~Rojas\orcid{0000-0003-1391-6854}\inst{\ref{aff2}}
\and M.~Melchior\inst{\ref{aff1}}
\and N.~E.~P.~Lines\orcid{0009-0004-7751-1914}\inst{\ref{aff3}}
\and T.~E.~Collett\orcid{0000-0001-5564-3140}\inst{\ref{aff3}}
\and A.~Verma\orcid{0000-0002-0730-0781}\inst{\ref{aff4}}
\and P.~Holloway\orcid{0009-0002-8896-6100}\inst{\ref{aff3}}
\and G.~Despali\orcid{0000-0001-6150-4112}\inst{\ref{aff5},\ref{aff6},\ref{aff7}}
\and S.~Schuldt\orcid{0000-0003-2497-6334}\inst{\ref{aff8},\ref{aff9}}
\and R.~B.~Metcalf\orcid{0000-0003-3167-2574}\inst{\ref{aff5},\ref{aff6}}
\and R.~Gavazzi\orcid{0000-0002-5540-6935}\inst{\ref{aff10},\ref{aff11}}
\and F.~Courbin\orcid{0000-0003-0758-6510}\inst{\ref{aff12},\ref{aff13},\ref{aff14}}
\and J.~A.~Acevedo~Barroso\orcid{0000-0002-9654-1711}\inst{\ref{aff15},\ref{aff16}}
\and B.~Cl\'ement\orcid{0000-0002-7966-3661}\inst{\ref{aff15},\ref{aff17}}
\and T.~Li\orcid{0009-0005-5008-0381}\inst{\ref{aff3}}
\and D.~Sluse\orcid{0000-0001-6116-2095}\inst{\ref{aff18}}
\and J.~Wilde\orcid{0000-0002-4460-7379}\inst{\ref{aff12}}
\and A.~Melo\orcid{0000-0002-6449-3970}\inst{\ref{aff19},\ref{aff20},\ref{aff21}}
\and A.~Sonnenfeld\orcid{0000-0002-6061-5977}\inst{\ref{aff22}}
\and C.~Tortora\orcid{0000-0001-7958-6531}\inst{\ref{aff23}}
\and T.~T.~Thai\orcid{0000-0002-8408-4816}\inst{\ref{aff24}}
\and M.~Millon\orcid{0000-0001-7051-497X}\inst{\ref{aff25},\ref{aff26}}
\and C.~Spiniello\orcid{0000-0002-3909-6359}\inst{\ref{aff4}}
\and A.~Manj\'on-Garc\'ia\orcid{0000-0002-7413-8825}\inst{\ref{aff27}}
\and M.~Meneghetti\orcid{0000-0003-1225-7084}\inst{\ref{aff6},\ref{aff7}}
\and B.~C.~Nagam\orcid{0000-0002-3724-7694}\inst{\ref{aff28},\ref{aff29}}
\and B.~Altieri\orcid{0000-0003-3936-0284}\inst{\ref{aff30}}
\and S.~Andreon\orcid{0000-0002-2041-8784}\inst{\ref{aff31}}
\and N.~Auricchio\orcid{0000-0003-4444-8651}\inst{\ref{aff6}}
\and C.~Baccigalupi\orcid{0000-0002-8211-1630}\inst{\ref{aff32},\ref{aff33},\ref{aff34},\ref{aff35}}
\and M.~Baldi\orcid{0000-0003-4145-1943}\inst{\ref{aff36},\ref{aff6},\ref{aff7}}
\and A.~Balestra\orcid{0000-0002-6967-261X}\inst{\ref{aff37}}
\and S.~Bardelli\orcid{0000-0002-8900-0298}\inst{\ref{aff6}}
\and P.~Battaglia\orcid{0000-0002-7337-5909}\inst{\ref{aff6}}
\and A.~Biviano\orcid{0000-0002-0857-0732}\inst{\ref{aff33},\ref{aff32}}
\and E.~Branchini\orcid{0000-0002-0808-6908}\inst{\ref{aff38},\ref{aff39},\ref{aff31}}
\and M.~Brescia\orcid{0000-0001-9506-5680}\inst{\ref{aff40},\ref{aff23}}
\and S.~Camera\orcid{0000-0003-3399-3574}\inst{\ref{aff41},\ref{aff42},\ref{aff43}}
\and V.~Capobianco\orcid{0000-0002-3309-7692}\inst{\ref{aff43}}
\and C.~Carbone\orcid{0000-0003-0125-3563}\inst{\ref{aff9}}
\and J.~Carretero\orcid{0000-0002-3130-0204}\inst{\ref{aff44},\ref{aff45}}
\and S.~Casas\orcid{0000-0002-4751-5138}\inst{\ref{aff46},\ref{aff47}}
\and M.~Castellano\orcid{0000-0001-9875-8263}\inst{\ref{aff48}}
\and G.~Castignani\orcid{0000-0001-6831-0687}\inst{\ref{aff6}}
\and S.~Cavuoti\orcid{0000-0002-3787-4196}\inst{\ref{aff23},\ref{aff49}}
\and A.~Cimatti\inst{\ref{aff50}}
\and C.~Colodro-Conde\inst{\ref{aff51}}
\and G.~Congedo\orcid{0000-0003-2508-0046}\inst{\ref{aff52}}
\and C.~J.~Conselice\orcid{0000-0003-1949-7638}\inst{\ref{aff53}}
\and L.~Conversi\orcid{0000-0002-6710-8476}\inst{\ref{aff54},\ref{aff30}}
\and Y.~Copin\orcid{0000-0002-5317-7518}\inst{\ref{aff55}}
\and A.~Costille\inst{\ref{aff10}}
\and H.~M.~Courtois\orcid{0000-0003-0509-1776}\inst{\ref{aff56}}
\and M.~Cropper\orcid{0000-0003-4571-9468}\inst{\ref{aff57}}
\and A.~Da~Silva\orcid{0000-0002-6385-1609}\inst{\ref{aff58},\ref{aff59}}
\and H.~Degaudenzi\orcid{0000-0002-5887-6799}\inst{\ref{aff60}}
\and G.~De~Lucia\orcid{0000-0002-6220-9104}\inst{\ref{aff33}}
\and C.~Dolding\orcid{0009-0003-7199-6108}\inst{\ref{aff57}}
\and H.~Dole\orcid{0000-0002-9767-3839}\inst{\ref{aff61}}
\and M.~Douspis\orcid{0000-0003-4203-3954}\inst{\ref{aff61}}
\and F.~Dubath\orcid{0000-0002-6533-2810}\inst{\ref{aff60}}
\and X.~Dupac\inst{\ref{aff30}}
\and S.~Dusini\orcid{0000-0002-1128-0664}\inst{\ref{aff62}}
\and S.~Escoffier\orcid{0000-0002-2847-7498}\inst{\ref{aff63}}
\and M.~Farina\orcid{0000-0002-3089-7846}\inst{\ref{aff64}}
\and F.~Faustini\orcid{0000-0001-6274-5145}\inst{\ref{aff48},\ref{aff65}}
\and S.~Ferriol\inst{\ref{aff55}}
\and F.~Finelli\orcid{0000-0002-6694-3269}\inst{\ref{aff6},\ref{aff66}}
\and P.~Fosalba\orcid{0000-0002-1510-5214}\inst{\ref{aff67},\ref{aff68}}
\and S.~Fotopoulou\orcid{0000-0002-9686-254X}\inst{\ref{aff69}}
\and M.~Frailis\orcid{0000-0002-7400-2135}\inst{\ref{aff33}}
\and E.~Franceschi\orcid{0000-0002-0585-6591}\inst{\ref{aff6}}
\and M.~Fumana\orcid{0000-0001-6787-5950}\inst{\ref{aff9}}
\and S.~Galeotta\orcid{0000-0002-3748-5115}\inst{\ref{aff33}}
\and K.~George\orcid{0000-0002-1734-8455}\inst{\ref{aff70}}
\and B.~Gillis\orcid{0000-0002-4478-1270}\inst{\ref{aff52}}
\and C.~Giocoli\orcid{0000-0002-9590-7961}\inst{\ref{aff6},\ref{aff7}}
\and J.~Gracia-Carpio\orcid{0000-0003-4689-3134}\inst{\ref{aff71}}
\and A.~Grazian\orcid{0000-0002-5688-0663}\inst{\ref{aff37}}
\and F.~Grupp\inst{\ref{aff71},\ref{aff72}}
\and L.~Guzzo\orcid{0000-0001-8264-5192}\inst{\ref{aff8},\ref{aff31},\ref{aff73}}
\and S.~V.~H.~Haugan\orcid{0000-0001-9648-7260}\inst{\ref{aff74}}
\and J.~Hoar\inst{\ref{aff30}}
\and W.~Holmes\inst{\ref{aff16}}
\and I.~M.~Hook\orcid{0000-0002-2960-978X}\inst{\ref{aff75}}
\and F.~Hormuth\inst{\ref{aff76}}
\and A.~Hornstrup\orcid{0000-0002-3363-0936}\inst{\ref{aff77},\ref{aff78}}
\and K.~Jahnke\orcid{0000-0003-3804-2137}\inst{\ref{aff79}}
\and M.~Jhabvala\inst{\ref{aff80}}
\and B.~Joachimi\orcid{0000-0001-7494-1303}\inst{\ref{aff81}}
\and S.~Kermiche\orcid{0000-0002-0302-5735}\inst{\ref{aff63}}
\and A.~Kiessling\orcid{0000-0002-2590-1273}\inst{\ref{aff16}}
\and B.~Kubik\orcid{0009-0006-5823-4880}\inst{\ref{aff55}}
\and M.~K\"ummel\orcid{0000-0003-2791-2117}\inst{\ref{aff72}}
\and M.~Kunz\orcid{0000-0002-3052-7394}\inst{\ref{aff26}}
\and H.~Kurki-Suonio\orcid{0000-0002-4618-3063}\inst{\ref{aff82},\ref{aff83}}
\and A.~M.~C.~Le~Brun\orcid{0000-0002-0936-4594}\inst{\ref{aff84}}
\and S.~Ligori\orcid{0000-0003-4172-4606}\inst{\ref{aff43}}
\and P.~B.~Lilje\orcid{0000-0003-4324-7794}\inst{\ref{aff74}}
\and V.~Lindholm\orcid{0000-0003-2317-5471}\inst{\ref{aff82},\ref{aff83}}
\and I.~Lloro\orcid{0000-0001-5966-1434}\inst{\ref{aff85}}
\and G.~Mainetti\orcid{0000-0003-2384-2377}\inst{\ref{aff86}}
\and E.~Maiorano\orcid{0000-0003-2593-4355}\inst{\ref{aff6}}
\and O.~Mansutti\orcid{0000-0001-5758-4658}\inst{\ref{aff33}}
\and S.~Marcin\inst{\ref{aff2}}
\and O.~Marggraf\orcid{0000-0001-7242-3852}\inst{\ref{aff87}}
\and M.~Martinelli\orcid{0000-0002-6943-7732}\inst{\ref{aff48},\ref{aff88}}
\and N.~Martinet\orcid{0000-0003-2786-7790}\inst{\ref{aff10}}
\and F.~Marulli\orcid{0000-0002-8850-0303}\inst{\ref{aff5},\ref{aff6},\ref{aff7}}
\and R.~J.~Massey\orcid{0000-0002-6085-3780}\inst{\ref{aff89}}
\and E.~Medinaceli\orcid{0000-0002-4040-7783}\inst{\ref{aff6}}
\and S.~Mei\orcid{0000-0002-2849-559X}\inst{\ref{aff90},\ref{aff91}}
\and E.~Merlin\orcid{0000-0001-6870-8900}\inst{\ref{aff48}}
\and G.~Meylan\inst{\ref{aff15}}
\and A.~Mora\orcid{0000-0002-1922-8529}\inst{\ref{aff92}}
\and M.~Moresco\orcid{0000-0002-7616-7136}\inst{\ref{aff5},\ref{aff6}}
\and L.~Moscardini\orcid{0000-0002-3473-6716}\inst{\ref{aff5},\ref{aff6},\ref{aff7}}
\and E.~Munari\orcid{0000-0002-1751-5946}\inst{\ref{aff33},\ref{aff32}}
\and R.~Nakajima\orcid{0009-0009-1213-7040}\inst{\ref{aff87}}
\and C.~Neissner\orcid{0000-0001-8524-4968}\inst{\ref{aff93},\ref{aff45}}
\and R.~C.~Nichol\orcid{0000-0003-0939-6518}\inst{\ref{aff94}}
\and S.-M.~Niemi\orcid{0009-0005-0247-0086}\inst{\ref{aff95}}
\and J.~W.~Nightingale\orcid{0000-0002-8987-7401}\inst{\ref{aff96}}
\and C.~Padilla\orcid{0000-0001-7951-0166}\inst{\ref{aff93}}
\and S.~Paltani\orcid{0000-0002-8108-9179}\inst{\ref{aff60}}
\and F.~Pasian\orcid{0000-0002-4869-3227}\inst{\ref{aff33}}
\and K.~Pedersen\inst{\ref{aff97}}
\and W.~J.~Percival\orcid{0000-0002-0644-5727}\inst{\ref{aff98},\ref{aff99},\ref{aff100}}
\and V.~Pettorino\orcid{0000-0002-4203-9320}\inst{\ref{aff95}}
\and S.~Pires\orcid{0000-0002-0249-2104}\inst{\ref{aff101}}
\and G.~Polenta\orcid{0000-0003-4067-9196}\inst{\ref{aff65}}
\and M.~Poncet\inst{\ref{aff102}}
\and L.~A.~Popa\inst{\ref{aff103}}
\and F.~Raison\orcid{0000-0002-7819-6918}\inst{\ref{aff71}}
\and A.~Renzi\orcid{0000-0001-9856-1970}\inst{\ref{aff104},\ref{aff62}}
\and J.~Rhodes\orcid{0000-0002-4485-8549}\inst{\ref{aff16}}
\and G.~Riccio\inst{\ref{aff23}}
\and E.~Romelli\orcid{0000-0003-3069-9222}\inst{\ref{aff33}}
\and M.~Roncarelli\orcid{0000-0001-9587-7822}\inst{\ref{aff6}}
\and R.~Saglia\orcid{0000-0003-0378-7032}\inst{\ref{aff72},\ref{aff71}}
\and Z.~Sakr\orcid{0000-0002-4823-3757}\inst{\ref{aff105},\ref{aff106},\ref{aff107}}
\and D.~Sapone\orcid{0000-0001-7089-4503}\inst{\ref{aff108}}
\and B.~Sartoris\orcid{0000-0003-1337-5269}\inst{\ref{aff72},\ref{aff33}}
\and M.~Schirmer\orcid{0000-0003-2568-9994}\inst{\ref{aff79}}
\and P.~Schneider\orcid{0000-0001-8561-2679}\inst{\ref{aff87}}
\and A.~Secroun\orcid{0000-0003-0505-3710}\inst{\ref{aff63}}
\and G.~Seidel\orcid{0000-0003-2907-353X}\inst{\ref{aff79}}
\and E.~Sihvola\orcid{0000-0003-1804-7715}\inst{\ref{aff109}}
\and P.~Simon\inst{\ref{aff87}}
\and C.~Sirignano\orcid{0000-0002-0995-7146}\inst{\ref{aff104},\ref{aff62}}
\and G.~Sirri\orcid{0000-0003-2626-2853}\inst{\ref{aff7}}
\and L.~Stanco\orcid{0000-0002-9706-5104}\inst{\ref{aff62}}
\and P.~Tallada-Cresp\'{i}\orcid{0000-0002-1336-8328}\inst{\ref{aff44},\ref{aff45}}
\and A.~N.~Taylor\inst{\ref{aff52}}
\and I.~Tereno\orcid{0000-0002-4537-6218}\inst{\ref{aff58},\ref{aff110}}
\and N.~Tessore\orcid{0000-0002-9696-7931}\inst{\ref{aff57}}
\and S.~Toft\orcid{0000-0003-3631-7176}\inst{\ref{aff111},\ref{aff112}}
\and F.~Torradeflot\orcid{0000-0003-1160-1517}\inst{\ref{aff45},\ref{aff44}}
\and I.~Tutusaus\orcid{0000-0002-3199-0399}\inst{\ref{aff68},\ref{aff67},\ref{aff105}}
\and L.~Valenziano\orcid{0000-0002-1170-0104}\inst{\ref{aff6},\ref{aff66}}
\and J.~Valiviita\orcid{0000-0001-6225-3693}\inst{\ref{aff82},\ref{aff83}}
\and T.~Vassallo\orcid{0000-0001-6512-6358}\inst{\ref{aff33},\ref{aff70}}
\and G.~Verdoes~Kleijn\orcid{0000-0001-5803-2580}\inst{\ref{aff29}}
\and A.~Veropalumbo\orcid{0000-0003-2387-1194}\inst{\ref{aff31},\ref{aff39},\ref{aff38}}
\and Y.~Wang\orcid{0000-0002-4749-2984}\inst{\ref{aff113}}
\and J.~Weller\orcid{0000-0002-8282-2010}\inst{\ref{aff72},\ref{aff71}}
\and A.~Zacchei\orcid{0000-0003-0396-1192}\inst{\ref{aff33},\ref{aff32}}
\and G.~Zamorani\orcid{0000-0002-2318-301X}\inst{\ref{aff6}}
\and E.~Zucca\orcid{0000-0002-5845-8132}\inst{\ref{aff6}}
\and M.~Ballardini\orcid{0000-0003-4481-3559}\inst{\ref{aff114},\ref{aff115},\ref{aff6}}
\and E.~Bozzo\orcid{0000-0002-8201-1525}\inst{\ref{aff60}}
\and C.~Burigana\orcid{0000-0002-3005-5796}\inst{\ref{aff116},\ref{aff66}}
\and R.~Cabanac\orcid{0000-0001-6679-2600}\inst{\ref{aff105}}
\and M.~Calabrese\orcid{0000-0002-2637-2422}\inst{\ref{aff117},\ref{aff9}}
\and A.~Cappi\inst{\ref{aff118},\ref{aff6}}
\and T.~Castro\orcid{0000-0002-6292-3228}\inst{\ref{aff33},\ref{aff34},\ref{aff32},\ref{aff119}}
\and J.~A.~Escartin~Vigo\inst{\ref{aff71}}
\and L.~Gabarra\orcid{0000-0002-8486-8856}\inst{\ref{aff4}}
\and S.~Hemmati\orcid{0000-0003-2226-5395}\inst{\ref{aff113}}
\and J.~Macias-Perez\orcid{0000-0002-5385-2763}\inst{\ref{aff120}}
\and R.~Maoli\orcid{0000-0002-6065-3025}\inst{\ref{aff121},\ref{aff48}}
\and J.~Mart\'{i}n-Fleitas\orcid{0000-0002-8594-569X}\inst{\ref{aff122}}
\and N.~Mauri\orcid{0000-0001-8196-1548}\inst{\ref{aff50},\ref{aff7}}
\and P.~Monaco\orcid{0000-0003-2083-7564}\inst{\ref{aff123},\ref{aff33},\ref{aff34},\ref{aff32}}
\and A.~A.~Nucita\inst{\ref{aff124},\ref{aff125},\ref{aff126}}
\and A.~Pezzotta\orcid{0000-0003-0726-2268}\inst{\ref{aff31}}
\and M.~P\"ontinen\orcid{0000-0001-5442-2530}\inst{\ref{aff82}}
\and I.~Risso\orcid{0000-0003-2525-7761}\inst{\ref{aff31},\ref{aff39}}
\and V.~Scottez\orcid{0009-0008-3864-940X}\inst{\ref{aff127},\ref{aff128}}
\and M.~Sereno\orcid{0000-0003-0302-0325}\inst{\ref{aff6},\ref{aff7}}
\and M.~Tenti\orcid{0000-0002-4254-5901}\inst{\ref{aff7}}
\and M.~Tucci\inst{\ref{aff60}}
\and M.~Viel\orcid{0000-0002-2642-5707}\inst{\ref{aff32},\ref{aff33},\ref{aff35},\ref{aff34},\ref{aff119}}
\and M.~Wiesmann\orcid{0009-0000-8199-5860}\inst{\ref{aff74}}
\and Y.~Akrami\orcid{0000-0002-2407-7956}\inst{\ref{aff107},\ref{aff129}}
\and I.~T.~Andika\orcid{0000-0001-6102-9526}\inst{\ref{aff70}}
\and G.~Angora\orcid{0000-0002-0316-6562}\inst{\ref{aff23},\ref{aff114}}
\and S.~Anselmi\orcid{0000-0002-3579-9583}\inst{\ref{aff62},\ref{aff104},\ref{aff130}}
\and M.~Archidiacono\orcid{0000-0003-4952-9012}\inst{\ref{aff8},\ref{aff73}}
\and F.~Atrio-Barandela\orcid{0000-0002-2130-2513}\inst{\ref{aff131}}
\and L.~Bazzanini\orcid{0000-0003-0727-0137}\inst{\ref{aff114},\ref{aff6}}
\and P.~Bergamini\orcid{0000-0003-1383-9414}\inst{\ref{aff6}}
\and D.~Bertacca\orcid{0000-0002-2490-7139}\inst{\ref{aff104},\ref{aff37},\ref{aff62}}
\and M.~Bethermin\orcid{0000-0002-3915-2015}\inst{\ref{aff132}}
\and F.~Beutler\orcid{0000-0003-0467-5438}\inst{\ref{aff52}}
\and A.~Blanchard\orcid{0000-0001-8555-9003}\inst{\ref{aff105}}
\and L.~Blot\orcid{0000-0002-9622-7167}\inst{\ref{aff133},\ref{aff84}}
\and S.~Borgani\orcid{0000-0001-6151-6439}\inst{\ref{aff123},\ref{aff32},\ref{aff33},\ref{aff34},\ref{aff119}}
\and M.~L.~Brown\orcid{0000-0002-0370-8077}\inst{\ref{aff53}}
\and S.~Bruton\orcid{0000-0002-6503-5218}\inst{\ref{aff134}}
\and A.~Calabro\orcid{0000-0003-2536-1614}\inst{\ref{aff48}}
\and B.~Camacho~Quevedo\orcid{0000-0002-8789-4232}\inst{\ref{aff32},\ref{aff35},\ref{aff33}}
\and F.~Caro\inst{\ref{aff48}}
\and C.~S.~Carvalho\inst{\ref{aff110}}
\and Y.~Charles\orcid{0009-0000-3636-834X}\inst{\ref{aff10}}
\and F.~Cogato\orcid{0000-0003-4632-6113}\inst{\ref{aff5},\ref{aff6}}
\and S.~Conseil\orcid{0000-0002-3657-4191}\inst{\ref{aff55}}
\and A.~R.~Cooray\orcid{0000-0002-3892-0190}\inst{\ref{aff135}}
\and O.~Cucciati\orcid{0000-0002-9336-7551}\inst{\ref{aff6}}
\and S.~Davini\orcid{0000-0003-3269-1718}\inst{\ref{aff39}}
\and F.~De~Paolis\orcid{0000-0001-6460-7563}\inst{\ref{aff124},\ref{aff125},\ref{aff126}}
\and G.~Desprez\orcid{0000-0001-8325-1742}\inst{\ref{aff29}}
\and A.~D\'iaz-S\'anchez\orcid{0000-0003-0748-4768}\inst{\ref{aff27}}
\and S.~Di~Domizio\orcid{0000-0003-2863-5895}\inst{\ref{aff38},\ref{aff39}}
\and J.~M.~Diego\orcid{0000-0001-9065-3926}\inst{\ref{aff136}}
\and P.-A.~Duc\orcid{0000-0003-3343-6284}\inst{\ref{aff132}}
\and V.~Duret\orcid{0009-0009-0383-4960}\inst{\ref{aff63}}
\and M.~Y.~Elkhashab\orcid{0000-0001-9306-2603}\inst{\ref{aff33},\ref{aff34},\ref{aff123},\ref{aff32}}
\and A.~Enia\orcid{0000-0002-0200-2857}\inst{\ref{aff6}}
\and Y.~Fang\orcid{0000-0002-0334-6950}\inst{\ref{aff72}}
\and A.~Finoguenov\orcid{0000-0002-4606-5403}\inst{\ref{aff82}}
\and A.~Franco\orcid{0000-0002-4761-366X}\inst{\ref{aff125},\ref{aff124},\ref{aff126}}
\and K.~Ganga\orcid{0000-0001-8159-8208}\inst{\ref{aff90}}
\and T.~Gasparetto\orcid{0000-0002-7913-4866}\inst{\ref{aff48}}
\and E.~Gaztanaga\orcid{0000-0001-9632-0815}\inst{\ref{aff68},\ref{aff67},\ref{aff3}}
\and F.~Giacomini\orcid{0000-0002-3129-2814}\inst{\ref{aff7}}
\and F.~Gianotti\orcid{0000-0003-4666-119X}\inst{\ref{aff6}}
\and G.~Gozaliasl\orcid{0000-0002-0236-919X}\inst{\ref{aff137},\ref{aff82}}
\and M.~Guidi\orcid{0000-0001-9408-1101}\inst{\ref{aff36},\ref{aff6}}
\and C.~M.~Gutierrez\orcid{0000-0001-7854-783X}\inst{\ref{aff51},\ref{aff138}}
\and A.~Hall\orcid{0000-0002-3139-8651}\inst{\ref{aff52}}
\and C.~Hern\'andez-Monteagudo\orcid{0000-0001-5471-9166}\inst{\ref{aff138},\ref{aff51}}
\and H.~Hildebrandt\orcid{0000-0002-9814-3338}\inst{\ref{aff139}}
\and J.~Hjorth\orcid{0000-0002-4571-2306}\inst{\ref{aff97}}
\and J.~J.~E.~Kajava\orcid{0000-0002-3010-8333}\inst{\ref{aff140},\ref{aff141},\ref{aff142}}
\and Y.~Kang\orcid{0009-0000-8588-7250}\inst{\ref{aff60}}
\and V.~Kansal\orcid{0000-0002-4008-6078}\inst{\ref{aff143},\ref{aff144}}
\and D.~Karagiannis\orcid{0000-0002-4927-0816}\inst{\ref{aff114},\ref{aff145}}
\and K.~Kiiveri\inst{\ref{aff109}}
\and J.~Kim\orcid{0000-0003-2776-2761}\inst{\ref{aff4}}
\and C.~C.~Kirkpatrick\inst{\ref{aff109}}
\and S.~Kruk\orcid{0000-0001-8010-8879}\inst{\ref{aff30}}
\and F.~Lepori\orcid{0009-0000-5061-7138}\inst{\ref{aff146}}
\and G.~Leroy\orcid{0009-0004-2523-4425}\inst{\ref{aff147},\ref{aff89}}
\and J.~Lesgourgues\orcid{0000-0001-7627-353X}\inst{\ref{aff46}}
\and T.~I.~Liaudat\orcid{0000-0002-9104-314X}\inst{\ref{aff148}}
\and S.~J.~Liu\orcid{0000-0001-7680-2139}\inst{\ref{aff64}}
\and A.~Loureiro\orcid{0000-0002-4371-0876}\inst{\ref{aff149},\ref{aff150}}
\and M.~Magliocchetti\orcid{0000-0001-9158-4838}\inst{\ref{aff64}}
\and E.~A.~Magnier\orcid{0000-0002-7965-2815}\inst{\ref{aff151}}
\and F.~Mannucci\orcid{0000-0002-4803-2381}\inst{\ref{aff152}}
\and C.~J.~A.~P.~Martins\orcid{0000-0002-4886-9261}\inst{\ref{aff153},\ref{aff154}}
\and L.~Maurin\orcid{0000-0002-8406-0857}\inst{\ref{aff61}}
\and M.~Miluzio\inst{\ref{aff30},\ref{aff155}}
\and C.~Moretti\orcid{0000-0003-3314-8936}\inst{\ref{aff33},\ref{aff32},\ref{aff34}}
\and G.~Morgante\inst{\ref{aff6}}
\and K.~Naidoo\orcid{0000-0002-9182-1802}\inst{\ref{aff3},\ref{aff79}}
\and A.~Navarro-Alsina\orcid{0000-0002-3173-2592}\inst{\ref{aff87}}
\and S.~Nesseris\orcid{0000-0002-0567-0324}\inst{\ref{aff107}}
\and D.~Paoletti\orcid{0000-0003-4761-6147}\inst{\ref{aff6},\ref{aff66}}
\and F.~Passalacqua\orcid{0000-0002-8606-4093}\inst{\ref{aff104},\ref{aff62}}
\and K.~Paterson\orcid{0000-0001-8340-3486}\inst{\ref{aff79}}
\and L.~Patrizii\inst{\ref{aff7}}
\and D.~Potter\orcid{0000-0002-0757-5195}\inst{\ref{aff156}}
\and G.~W.~Pratt\inst{\ref{aff101}}
\and S.~Quai\orcid{0000-0002-0449-8163}\inst{\ref{aff5},\ref{aff6}}
\and M.~Radovich\orcid{0000-0002-3585-866X}\inst{\ref{aff37}}
\and W.~Roster\orcid{0000-0002-9149-6528}\inst{\ref{aff71}}
\and S.~Sacquegna\orcid{0000-0002-8433-6630}\inst{\ref{aff157}}
\and M.~Sahl\'en\orcid{0000-0003-0973-4804}\inst{\ref{aff158}}
\and D.~B.~Sanders\orcid{0000-0002-1233-9998}\inst{\ref{aff151}}
\and E.~Sarpa\orcid{0000-0002-1256-655X}\inst{\ref{aff35},\ref{aff119},\ref{aff33}}
\and C.~Scarlata\orcid{0000-0002-9136-8876}\inst{\ref{aff28}}
\and A.~Schneider\orcid{0000-0001-7055-8104}\inst{\ref{aff156}}
\and M.~Schultheis\inst{\ref{aff118}}
\and D.~Sciotti\orcid{0009-0008-4519-2620}\inst{\ref{aff48},\ref{aff88}}
\and E.~Sellentin\inst{\ref{aff159},\ref{aff160}}
\and L.~C.~Smith\orcid{0000-0002-3259-2771}\inst{\ref{aff161}}
\and J.~G.~Sorce\orcid{0000-0002-2307-2432}\inst{\ref{aff162},\ref{aff61}}
\and K.~Tanidis\orcid{0000-0001-9843-5130}\inst{\ref{aff163}}
\and C.~Tao\orcid{0000-0001-7961-8177}\inst{\ref{aff63}}
\and F.~Tarsitano\orcid{0000-0002-5919-0238}\inst{\ref{aff25},\ref{aff60}}
\and G.~Testera\inst{\ref{aff39}}
\and R.~Teyssier\orcid{0000-0001-7689-0933}\inst{\ref{aff164}}
\and S.~Tosi\orcid{0000-0002-7275-9193}\inst{\ref{aff38},\ref{aff31},\ref{aff39}}
\and A.~Troja\orcid{0000-0003-0239-4595}\inst{\ref{aff104},\ref{aff62}}
\and A.~Venhola\orcid{0000-0001-6071-4564}\inst{\ref{aff165}}
\and D.~Vergani\orcid{0000-0003-0898-2216}\inst{\ref{aff6}}
\and G.~Vernardos\orcid{0000-0001-8554-7248}\inst{\ref{aff166},\ref{aff167}}
\and G.~Verza\orcid{0000-0002-1886-8348}\inst{\ref{aff168},\ref{aff169}}
\and S.~Vinciguerra\orcid{0009-0005-4018-3184}\inst{\ref{aff10}}
\and M.~Walmsley\orcid{0000-0002-6408-4181}\inst{\ref{aff170},\ref{aff53}}
\and N.~A.~Walton\orcid{0000-0003-3983-8778}\inst{\ref{aff161}}
\and A.~H.~Wright\orcid{0000-0001-7363-7932}\inst{\ref{aff139}}}
										   
\institute{University of Applied Sciences and Arts of Northwestern Switzerland, School of Engineering, 5210 Windisch, Switzerland\label{aff1}
\and
University of Applied Sciences and Arts of Northwestern Switzerland, School of Computer Science, 5210 Windisch, Switzerland\label{aff2}
\and
Institute of Cosmology and Gravitation, University of Portsmouth, Portsmouth PO1 3FX, UK\label{aff3}
\and
Department of Physics, Oxford University, Keble Road, Oxford OX1 3RH, UK\label{aff4}
\and
Dipartimento di Fisica e Astronomia "Augusto Righi" - Alma Mater Studiorum Universit\`a di Bologna, via Piero Gobetti 93/2, 40129 Bologna, Italy\label{aff5}
\and
INAF-Osservatorio di Astrofisica e Scienza dello Spazio di Bologna, Via Piero Gobetti 93/3, 40129 Bologna, Italy\label{aff6}
\and
INFN-Sezione di Bologna, Viale Berti Pichat 6/2, 40127 Bologna, Italy\label{aff7}
\and
Dipartimento di Fisica "Aldo Pontremoli", Universit\`a degli Studi di Milano, Via Celoria 16, 20133 Milano, Italy\label{aff8}
\and
INAF-IASF Milano, Via Alfonso Corti 12, 20133 Milano, Italy\label{aff9}
\and
Aix-Marseille Universit\'e, CNRS, CNES, LAM, Marseille, France\label{aff10}
\and
Institut d'Astrophysique de Paris, UMR 7095, CNRS, and Sorbonne Universit\'e, 98 bis boulevard Arago, 75014 Paris, France\label{aff11}
\and
Institut de Ci\`{e}ncies del Cosmos (ICCUB), Universitat de Barcelona (IEEC-UB), Mart\'{i} i Franqu\`{e}s 1, 08028 Barcelona, Spain\label{aff12}
\and
Instituci\'o Catalana de Recerca i Estudis Avan\c{c}ats (ICREA), Passeig de Llu\'{\i}s Companys 23, 08010 Barcelona, Spain\label{aff13}
\and
Institut de Ciencies de l'Espai (IEEC-CSIC), Campus UAB, Carrer de Can Magrans, s/n Cerdanyola del Vall\'es, 08193 Barcelona, Spain\label{aff14}
\and
Institute of Physics, Laboratory of Astrophysics, Ecole Polytechnique F\'ed\'erale de Lausanne (EPFL), Observatoire de Sauverny, 1290 Versoix, Switzerland\label{aff15}
\and
Jet Propulsion Laboratory, California Institute of Technology, 4800 Oak Grove Drive, Pasadena, CA, 91109, USA\label{aff16}
\and
SCITAS, Ecole Polytechnique F\'ed\'erale de Lausanne (EPFL), 1015 Lausanne, Switzerland\label{aff17}
\and
STAR Institute, University of Li{\`e}ge, Quartier Agora, All\'ee du six Ao\^ut 19c, 4000 Li\`ege, Belgium\label{aff18}
\and
Max-Planck-Institut f\"ur Astrophysik, Karl-Schwarzschild-Str.~1, 85748 Garching, Germany\label{aff19}
\and
Technical University of Munich, TUM School of Natural Sciences, Physics Department, James-Franck-Str.~1, 85748 Garching, Germany\label{aff20}
\and
European Southern Observatory, Karl-Schwarzschild-Str.~2, 85748 Garching, Germany\label{aff21}
\and
Department of Astronomy, School of Physics and Astronomy, Shanghai Jiao Tong University, Shanghai 200240, China\label{aff22}
\and
INAF-Osservatorio Astronomico di Capodimonte, Via Moiariello 16, 80131 Napoli, Italy\label{aff23}
\and
National Astronomical Observatory of Japan, 2-21-1 Osawa, Mitaka, Tokyo 181-8588, Japan\label{aff24}
\and
Institute for Particle Physics and Astrophysics, Dept. of Physics, ETH Zurich, Wolfgang-Pauli-Strasse 27, 8093 Zurich, Switzerland\label{aff25}
\and
Universit\'e de Gen\`eve, D\'epartement de Physique Th\'eorique and Centre for Astroparticle Physics, 24 quai Ernest-Ansermet, CH-1211 Gen\`eve 4, Switzerland\label{aff26}
\and
Departamento F\'isica Aplicada, Universidad Polit\'ecnica de Cartagena, Campus Muralla del Mar, 30202 Cartagena, Murcia, Spain\label{aff27}
\and
Minnesota Institute for Astrophysics, University of Minnesota, 116 Church St SE, Minneapolis, MN 55455, USA\label{aff28}
\and
Kapteyn Astronomical Institute, University of Groningen, PO Box 800, 9700 AV Groningen, The Netherlands\label{aff29}
\and
ESAC/ESA, Camino Bajo del Castillo, s/n., Urb. Villafranca del Castillo, 28692 Villanueva de la Ca\~nada, Madrid, Spain\label{aff30}
\and
INAF-Osservatorio Astronomico di Brera, Via Brera 28, 20122 Milano, Italy\label{aff31}
\and
IFPU, Institute for Fundamental Physics of the Universe, via Beirut 2, 34151 Trieste, Italy\label{aff32}
\and
INAF-Osservatorio Astronomico di Trieste, Via G. B. Tiepolo 11, 34143 Trieste, Italy\label{aff33}
\and
INFN, Sezione di Trieste, Via Valerio 2, 34127 Trieste TS, Italy\label{aff34}
\and
SISSA, International School for Advanced Studies, Via Bonomea 265, 34136 Trieste TS, Italy\label{aff35}
\and
Dipartimento di Fisica e Astronomia, Universit\`a di Bologna, Via Gobetti 93/2, 40129 Bologna, Italy\label{aff36}
\and
INAF-Osservatorio Astronomico di Padova, Via dell'Osservatorio 5, 35122 Padova, Italy\label{aff37}
\and
Dipartimento di Fisica, Universit\`a di Genova, Via Dodecaneso 33, 16146, Genova, Italy\label{aff38}
\and
INFN-Sezione di Genova, Via Dodecaneso 33, 16146, Genova, Italy\label{aff39}
\and
Department of Physics "E. Pancini", University Federico II, Via Cinthia 6, 80126, Napoli, Italy\label{aff40}
\and
Dipartimento di Fisica, Universit\`a degli Studi di Torino, Via P. Giuria 1, 10125 Torino, Italy\label{aff41}
\and
INFN-Sezione di Torino, Via P. Giuria 1, 10125 Torino, Italy\label{aff42}
\and
INAF-Osservatorio Astrofisico di Torino, Via Osservatorio 20, 10025 Pino Torinese (TO), Italy\label{aff43}
\and
Centro de Investigaciones Energ\'eticas, Medioambientales y Tecnol\'ogicas (CIEMAT), Avenida Complutense 40, 28040 Madrid, Spain\label{aff44}
\and
Port d'Informaci\'{o} Cient\'{i}fica, Campus UAB, C. Albareda s/n, 08193 Bellaterra (Barcelona), Spain\label{aff45}
\and
Institute for Theoretical Particle Physics and Cosmology (TTK), RWTH Aachen University, 52056 Aachen, Germany\label{aff46}
\and
Deutsches Zentrum f\"ur Luft- und Raumfahrt e. V. (DLR), Linder H\"ohe, 51147 K\"oln, Germany\label{aff47}
\and
INAF-Osservatorio Astronomico di Roma, Via Frascati 33, 00078 Monteporzio Catone, Italy\label{aff48}
\and
INFN section of Naples, Via Cinthia 6, 80126, Napoli, Italy\label{aff49}
\and
Dipartimento di Fisica e Astronomia "Augusto Righi" - Alma Mater Studiorum Universit\`a di Bologna, Viale Berti Pichat 6/2, 40127 Bologna, Italy\label{aff50}
\and
Instituto de Astrof\'{\i}sica de Canarias, E-38205 La Laguna, Tenerife, Spain\label{aff51}
\and
Institute for Astronomy, University of Edinburgh, Royal Observatory, Blackford Hill, Edinburgh EH9 3HJ, UK\label{aff52}
\and
Jodrell Bank Centre for Astrophysics, Department of Physics and Astronomy, University of Manchester, Oxford Road, Manchester M13 9PL, UK\label{aff53}
\and
European Space Agency/ESRIN, Largo Galileo Galilei 1, 00044 Frascati, Roma, Italy\label{aff54}
\and
Universit\'e Claude Bernard Lyon 1, CNRS/IN2P3, IP2I Lyon, UMR 5822, Villeurbanne, F-69100, France\label{aff55}
\and
UCB Lyon 1, CNRS/IN2P3, IUF, IP2I Lyon, 4 rue Enrico Fermi, 69622 Villeurbanne, France\label{aff56}
\and
Mullard Space Science Laboratory, University College London, Holmbury St Mary, Dorking, Surrey RH5 6NT, UK\label{aff57}
\and
Departamento de F\'isica, Faculdade de Ci\^encias, Universidade de Lisboa, Edif\'icio C8, Campo Grande, PT1749-016 Lisboa, Portugal\label{aff58}
\and
Instituto de Astrof\'isica e Ci\^encias do Espa\c{c}o, Faculdade de Ci\^encias, Universidade de Lisboa, Campo Grande, 1749-016 Lisboa, Portugal\label{aff59}
\and
Department of Astronomy, University of Geneva, ch. d'Ecogia 16, 1290 Versoix, Switzerland\label{aff60}
\and
Universit\'e Paris-Saclay, CNRS, Institut d'astrophysique spatiale, 91405, Orsay, France\label{aff61}
\and
INFN-Padova, Via Marzolo 8, 35131 Padova, Italy\label{aff62}
\and
Aix-Marseille Universit\'e, CNRS/IN2P3, CPPM, Marseille, France\label{aff63}
\and
INAF-Istituto di Astrofisica e Planetologia Spaziali, via del Fosso del Cavaliere, 100, 00100 Roma, Italy\label{aff64}
\and
Space Science Data Center, Italian Space Agency, via del Politecnico snc, 00133 Roma, Italy\label{aff65}
\and
INFN-Bologna, Via Irnerio 46, 40126 Bologna, Italy\label{aff66}
\and
Institut d'Estudis Espacials de Catalunya (IEEC),  Edifici RDIT, Campus UPC, 08860 Castelldefels, Barcelona, Spain\label{aff67}
\and
Institute of Space Sciences (ICE, CSIC), Campus UAB, Carrer de Can Magrans, s/n, 08193 Barcelona, Spain\label{aff68}
\and
School of Physics, HH Wills Physics Laboratory, University of Bristol, Tyndall Avenue, Bristol, BS8 1TL, UK\label{aff69}
\and
University Observatory, LMU Faculty of Physics, Scheinerstr.~1, 81679 Munich, Germany\label{aff70}
\and
Max Planck Institute for Extraterrestrial Physics, Giessenbachstr. 1, 85748 Garching, Germany\label{aff71}
\and
Universit\"ats-Sternwarte M\"unchen, Fakult\"at f\"ur Physik, Ludwig-Maximilians-Universit\"at M\"unchen, Scheinerstr.~1, 81679 M\"unchen, Germany\label{aff72}
\and
INFN-Sezione di Milano, Via Celoria 16, 20133 Milano, Italy\label{aff73}
\and
Institute of Theoretical Astrophysics, University of Oslo, P.O. Box 1029 Blindern, 0315 Oslo, Norway\label{aff74}
\and
Department of Physics, Lancaster University, Lancaster, LA1 4YB, UK\label{aff75}
\and
Felix Hormuth Engineering, Goethestr. 17, 69181 Leimen, Germany\label{aff76}
\and
Technical University of Denmark, Elektrovej 327, 2800 Kgs. Lyngby, Denmark\label{aff77}
\and
Cosmic Dawn Center (DAWN), Denmark\label{aff78}
\and
Max-Planck-Institut f\"ur Astronomie, K\"onigstuhl 17, 69117 Heidelberg, Germany\label{aff79}
\and
NASA Goddard Space Flight Center, Greenbelt, MD 20771, USA\label{aff80}
\and
Department of Physics and Astronomy, University College London, Gower Street, London WC1E 6BT, UK\label{aff81}
\and
Department of Physics, P.O. Box 64, University of Helsinki, 00014 Helsinki, Finland\label{aff82}
\and
Helsinki Institute of Physics, Gustaf H{\"a}llstr{\"o}min katu 2, University of Helsinki, 00014 Helsinki, Finland\label{aff83}
\and
Laboratoire d'etude de l'Univers et des phenomenes eXtremes, Observatoire de Paris, Universit\'e PSL, Sorbonne Universit\'e, CNRS, 92190 Meudon, France\label{aff84}
\and
SKAO, Jodrell Bank, Lower Withington, Macclesfield SK11 9FT, UK\label{aff85}
\and
Centre de Calcul de l'IN2P3/CNRS, 21 avenue Pierre de Coubertin 69627 Villeurbanne Cedex, France\label{aff86}
\and
Universit\"at Bonn, Argelander-Institut f\"ur Astronomie, Auf dem H\"ugel 71, 53121 Bonn, Germany\label{aff87}
\and
INFN-Sezione di Roma, Piazzale Aldo Moro, 2 - c/o Dipartimento di Fisica, Edificio G. Marconi, 00185 Roma, Italy\label{aff88}
\and
Department of Physics, Institute for Computational Cosmology, Durham University, South Road, Durham, DH1 3LE, UK\label{aff89}
\and
Universit\'e Paris Cit\'e, CNRS, Astroparticule et Cosmologie, 75013 Paris, France\label{aff90}
\and
CNRS-UCB International Research Laboratory, Centre Pierre Bin\'etruy, IRL2007, CPB-IN2P3, Berkeley, USA\label{aff91}
\and
Telespazio UK S.L. for European Space Agency (ESA), Camino bajo del Castillo, s/n, Urbanizacion Villafranca del Castillo, Villanueva de la Ca\~nada, 28692 Madrid, Spain\label{aff92}
\and
Institut de F\'{i}sica d'Altes Energies (IFAE), The Barcelona Institute of Science and Technology, Campus UAB, 08193 Bellaterra (Barcelona), Spain\label{aff93}
\and
School of Mathematics and Physics, University of Surrey, Guildford, Surrey, GU2 7XH, UK\label{aff94}
\and
European Space Agency/ESTEC, Keplerlaan 1, 2201 AZ Noordwijk, The Netherlands\label{aff95}
\and
School of Mathematics, Statistics and Physics, Newcastle University, Herschel Building, Newcastle-upon-Tyne, NE1 7RU, UK\label{aff96}
\and
DARK, Niels Bohr Institute, University of Copenhagen, Jagtvej 155, 2200 Copenhagen, Denmark\label{aff97}
\and
Waterloo Centre for Astrophysics, University of Waterloo, Waterloo, Ontario N2L 3G1, Canada\label{aff98}
\and
Department of Physics and Astronomy, University of Waterloo, Waterloo, Ontario N2L 3G1, Canada\label{aff99}
\and
Perimeter Institute for Theoretical Physics, Waterloo, Ontario N2L 2Y5, Canada\label{aff100}
\and
Universit\'e Paris-Saclay, Universit\'e Paris Cit\'e, CEA, CNRS, AIM, 91191, Gif-sur-Yvette, France\label{aff101}
\and
Centre National d'Etudes Spatiales -- Centre spatial de Toulouse, 18 avenue Edouard Belin, 31401 Toulouse Cedex 9, France\label{aff102}
\and
Institute of Space Science, Str. Atomistilor, nr. 409 M\u{a}gurele, Ilfov, 077125, Romania\label{aff103}
\and
Dipartimento di Fisica e Astronomia "G. Galilei", Universit\`a di Padova, Via Marzolo 8, 35131 Padova, Italy\label{aff104}
\and
Institut de Recherche en Astrophysique et Plan\'etologie (IRAP), Universit\'e de Toulouse, CNRS, UPS, CNES, 14 Av. Edouard Belin, 31400 Toulouse, France\label{aff105}
\and
Universit\'e St Joseph; Faculty of Sciences, Beirut, Lebanon\label{aff106}
\and
Instituto de F\'isica Te\'orica UAM-CSIC, Campus de Cantoblanco, 28049 Madrid, Spain\label{aff107}
\and
Departamento de F\'isica, FCFM, Universidad de Chile, Blanco Encalada 2008, Santiago, Chile\label{aff108}
\and
Department of Physics and Helsinki Institute of Physics, Gustaf H\"allstr\"omin katu 2, University of Helsinki, 00014 Helsinki, Finland\label{aff109}
\and
Instituto de Astrof\'isica e Ci\^encias do Espa\c{c}o, Faculdade de Ci\^encias, Universidade de Lisboa, Tapada da Ajuda, 1349-018 Lisboa, Portugal\label{aff110}
\and
Cosmic Dawn Center (DAWN)\label{aff111}
\and
Niels Bohr Institute, University of Copenhagen, Jagtvej 128, 2200 Copenhagen, Denmark\label{aff112}
\and
Caltech/IPAC, 1200 E. California Blvd., Pasadena, CA 91125, USA\label{aff113}
\and
Dipartimento di Fisica e Scienze della Terra, Universit\`a degli Studi di Ferrara, Via Giuseppe Saragat 1, 44122 Ferrara, Italy\label{aff114}
\and
Istituto Nazionale di Fisica Nucleare, Sezione di Ferrara, Via Giuseppe Saragat 1, 44122 Ferrara, Italy\label{aff115}
\and
INAF, Istituto di Radioastronomia, Via Piero Gobetti 101, 40129 Bologna, Italy\label{aff116}
\and
Astronomical Observatory of the Autonomous Region of the Aosta Valley (OAVdA), Loc. Lignan 39, I-11020, Nus (Aosta Valley), Italy\label{aff117}
\and
Universit\'e C\^{o}te d'Azur, Observatoire de la C\^{o}te d'Azur, CNRS, Laboratoire Lagrange, Bd de l'Observatoire, CS 34229, 06304 Nice cedex 4, France\label{aff118}
\and
ICSC - Centro Nazionale di Ricerca in High Performance Computing, Big Data e Quantum Computing, Via Magnanelli 2, Bologna, Italy\label{aff119}
\and
Univ. Grenoble Alpes, CNRS, Grenoble INP, LPSC-IN2P3, 53, Avenue des Martyrs, 38000, Grenoble, France\label{aff120}
\and
Dipartimento di Fisica, Sapienza Universit\`a di Roma, Piazzale Aldo Moro 2, 00185 Roma, Italy\label{aff121}
\and
Aurora Technology for European Space Agency (ESA), Camino bajo del Castillo, s/n, Urbanizacion Villafranca del Castillo, Villanueva de la Ca\~nada, 28692 Madrid, Spain\label{aff122}
\and
Dipartimento di Fisica - Sezione di Astronomia, Universit\`a di Trieste, Via Tiepolo 11, 34131 Trieste, Italy\label{aff123}
\and
Department of Mathematics and Physics E. De Giorgi, University of Salento, Via per Arnesano, CP-I93, 73100, Lecce, Italy\label{aff124}
\and
INFN, Sezione di Lecce, Via per Arnesano, CP-193, 73100, Lecce, Italy\label{aff125}
\and
INAF-Sezione di Lecce, c/o Dipartimento Matematica e Fisica, Via per Arnesano, 73100, Lecce, Italy\label{aff126}
\and
Institut d'Astrophysique de Paris, 98bis Boulevard Arago, 75014, Paris, France\label{aff127}
\and
ICL, Junia, Universit\'e Catholique de Lille, LITL, 59000 Lille, France\label{aff128}
\and
CERCA/ISO, Department of Physics, Case Western Reserve University, 10900 Euclid Avenue, Cleveland, OH 44106, USA\label{aff129}
\and
Laboratoire Univers et Th\'eorie, Observatoire de Paris, Universit\'e PSL, Universit\'e Paris Cit\'e, CNRS, 92190 Meudon, France\label{aff130}
\and
Departamento de F{\'\i}sica Fundamental. Universidad de Salamanca. Plaza de la Merced s/n. 37008 Salamanca, Spain\label{aff131}
\and
Universit\'e de Strasbourg, CNRS, Observatoire astronomique de Strasbourg, UMR 7550, 67000 Strasbourg, France\label{aff132}
\and
Center for Data-Driven Discovery, Kavli IPMU (WPI), UTIAS, The University of Tokyo, Kashiwa, Chiba 277-8583, Japan\label{aff133}
\and
California Institute of Technology, 1200 E California Blvd, Pasadena, CA 91125, USA\label{aff134}
\and
Department of Physics \& Astronomy, University of California Irvine, Irvine CA 92697, USA\label{aff135}
\and
Instituto de F\'isica de Cantabria, Edificio Juan Jord\'a, Avenida de los Castros, 39005 Santander, Spain\label{aff136}
\and
Department of Computer Science, Aalto University, PO Box 15400, Espoo, FI-00 076, Finland\label{aff137}
\and
Universidad de La Laguna, Dpto. Astrof\'\i sica, E-38206 La Laguna, Tenerife, Spain\label{aff138}
\and
Ruhr University Bochum, Faculty of Physics and Astronomy, Astronomical Institute (AIRUB), German Centre for Cosmological Lensing (GCCL), 44780 Bochum, Germany\label{aff139}
\and
Department of Physics and Astronomy, Vesilinnantie 5, University of Turku, 20014 Turku, Finland\label{aff140}
\and
Finnish Centre for Astronomy with ESO (FINCA), Quantum, Vesilinnantie 5, University of Turku, 20014 Turku, Finland\label{aff141}
\and
Serco for European Space Agency (ESA), Camino bajo del Castillo, s/n, Urbanizacion Villafranca del Castillo, Villanueva de la Ca\~nada, 28692 Madrid, Spain\label{aff142}
\and
ARC Centre of Excellence for Dark Matter Particle Physics, Melbourne, Australia\label{aff143}
\and
Centre for Astrophysics \& Supercomputing, Swinburne University of Technology,  Hawthorn, Victoria 3122, Australia\label{aff144}
\and
Department of Physics and Astronomy, University of the Western Cape, Bellville, Cape Town, 7535, South Africa\label{aff145}
\and
Departement of Theoretical Physics, University of Geneva, Switzerland\label{aff146}
\and
Department of Physics, Centre for Extragalactic Astronomy, Durham University, South Road, Durham, DH1 3LE, UK\label{aff147}
\and
IRFU, CEA, Universit\'e Paris-Saclay 91191 Gif-sur-Yvette Cedex, France\label{aff148}
\and
Oskar Klein Centre for Cosmoparticle Physics, Department of Physics, Stockholm University, Stockholm, SE-106 91, Sweden\label{aff149}
\and
Astrophysics Group, Blackett Laboratory, Imperial College London, London SW7 2AZ, UK\label{aff150}
\and
Institute for Astronomy, University of Hawaii, 2680 Woodlawn Drive, Honolulu, HI 96822, USA\label{aff151}
\and
INAF-Osservatorio Astrofisico di Arcetri, Largo E. Fermi 5, 50125, Firenze, Italy\label{aff152}
\and
Centro de Astrof\'{\i}sica da Universidade do Porto, Rua das Estrelas, 4150-762 Porto, Portugal\label{aff153}
\and
Instituto de Astrof\'isica e Ci\^encias do Espa\c{c}o, Universidade do Porto, CAUP, Rua das Estrelas, PT4150-762 Porto, Portugal\label{aff154}
\and
HE Space for European Space Agency (ESA), Camino bajo del Castillo, s/n, Urbanizacion Villafranca del Castillo, Villanueva de la Ca\~nada, 28692 Madrid, Spain\label{aff155}
\and
Department of Astrophysics, University of Zurich, Winterthurerstrasse 190, 8057 Zurich, Switzerland\label{aff156}
\and
INAF - Osservatorio Astronomico d'Abruzzo, Via Maggini, 64100, Teramo, Italy\label{aff157}
\and
Theoretical astrophysics, Department of Physics and Astronomy, Uppsala University, Box 516, 751 37 Uppsala, Sweden\label{aff158}
\and
Mathematical Institute, University of Leiden, Einsteinweg 55, 2333 CA Leiden, The Netherlands\label{aff159}
\and
Leiden Observatory, Leiden University, Einsteinweg 55, 2333 CC Leiden, The Netherlands\label{aff160}
\and
Institute of Astronomy, University of Cambridge, Madingley Road, Cambridge CB3 0HA, UK\label{aff161}
\and
Univ. Lille, CNRS, Centrale Lille, UMR 9189 CRIStAL, 59000 Lille, France\label{aff162}
\and
Center for Astrophysics and Cosmology, University of Nova Gorica, Nova Gorica, Slovenia\label{aff163}
\and
Department of Astrophysical Sciences, Peyton Hall, Princeton University, Princeton, NJ 08544, USA\label{aff164}
\and
Space physics and astronomy research unit, University of Oulu, Pentti Kaiteran katu 1, FI-90014 Oulu, Finland\label{aff165}
\and
Department of Physics and Astronomy, Lehman College of the CUNY, Bronx, NY 10468, USA\label{aff166}
\and
American Museum of Natural History, Department of Astrophysics, New York, NY 10024, USA\label{aff167}
\and
International Centre for Theoretical Physics (ICTP), Strada Costiera 11, 34151 Trieste, Italy\label{aff168}
\and
Center for Computational Astrophysics, Flatiron Institute, 162 5th Avenue, 10010, New York, NY, USA\label{aff169}
\and
David A. Dunlap Department of Astronomy \& Astrophysics, University of Toronto, 50 St George Street, Toronto, Ontario M5S 3H4, Canada\label{aff170}}    